\documentclass[fleqn,usenatbib]{mnras}

\usepackage{newtxtext,newtxmath}

\usepackage[T1]{fontenc}

\DeclareRobustCommand{\VAN}[3]{#2}
\let\VANthebibliography\thebibliography
\def\thebibliography{\DeclareRobustCommand{\VAN}[3]{##3}\VANthebibliography}

\usepackage{graphicx}	
\usepackage{amsmath}	
\usepackage{bm}  
\usepackage{ulem}
\usepackage{soul,xcolor}

\graphicspath{ {./figure/} }

\newcommand\pc{\mbox{{$\rm pc$}}}
\newcommand\kpc{\mbox{{$\rm kpc$}}}
\newcommand\K{\mbox{{$\rm K$}}}
\newcommand\Msun{\mbox{$\rm M_{\sun}$}}
\newcommand\cc{\mbox{$\rm cm^{-3}$}}
\newcommand\yr{\mbox{$\rm yr$}}
\newcommand\myr{\mbox{$\rm Myr$}}
\newcommand\gyr{\mbox{$\rm Gyr$}}
\newcommand\kms{\mbox{$\rm km\,s^{-1}$}}
\newcommand\sfrunit{\mbox{$\rm M_{\sun}\,yr^{-1}$}}

\newcommand\HI{\mbox{H\,{\scriptsize \sc I}}}
\newcommand\HII{\mbox{H\,{\scriptsize \sc II}}}

\title[CCCs and star formation in galaxy simulations]{Cloud-cloud collisions triggering star formation in galaxy simulations}

\author[S. Horie, T. Okamoto and A. Habe]{
    Shu Horie,$^{1}$\thanks{E-mail: \href{mailto:horie@astro1.sci.hokudai.ac.jp}{horie@astro1.sci.hokudai.ac.jp}}
    Takashi Okamoto$^{2}$ and 
    Asao Habe$^{2}$
    \\
    $^{1}$Department of Cosmosciences, Graduate School of Science, Hokkaido University, N10 W8, Kitaku, Sapporo 060-0810, Japan\\
    $^{2}$Department of Physics, Faculty of Science, Hokkaido University, N10 W8, Kitaku, Sapporo 060-0810, Japan\\
}

\date{Accepted XXX. Received YYY; in original form ZZZ}

\pubyear{2023}

\begin{document}
\label{firstpage}
\pagerange{\pageref{firstpage}--\pageref{lastpage}}
\maketitle

\begin{abstract}
Cloud-cloud collisions (CCCs) are expected to compress gas and trigger star formation.
However, it is not well understood how the collisions and the induced star formation affect galactic-scale properties.
By developing an on-the-fly algorithm to identify CCCs at each timestep in a galaxy simulation and a model that relates CCC-triggered star formation to collision speeds, we perform simulations of isolated galaxies to study the evolution of galaxies and giant molecular clouds (GMCs) with prescriptions of self-consistent CCC-driven star formation and stellar feedback.
We find that the simulation with the CCC-triggered star formation produces slightly higher star formation rates and a steeper Kennicutt-Schmidt relation than that with a more standard star formation recipe, although collision speeds and frequencies are insensitive to the star formation models.
In the simulation with the CCC model, about $70$ per cent of the stars are born via CCCs, and colliding GMCs with masses of $\approx 10^{5.5}\,\Msun$ are the main drivers of CCC-driven star formation.
In the simulation with the standard star formation recipe, about 50 per cent of stars are born in colliding GMCs even without the CCC-triggered star formation model.
These results suggest that CCCs may be one of the most important star formation processes in galaxy evolution.
Furthermore, we find that a post-processing analysis of CCCs, as used in previous studies in galaxy simulations, may lead to slightly greater collision speeds and significantly lower collision frequencies than the on-the-fly analysis.

\end{abstract}




\section{Introduction}
\label{intro}

Cloud-cloud collisions (CCCs) are expected to be a process that not only grows clouds but also efficiently compresses and triggers star formation.
Since star formation is one of the most fundamental processes in galaxy formation and evolution, CCCs may play an essential role on them.
However, there is still a lack of understanding of how CCCs affect star formation activity in galaxies, how CCC-driven star formation affects properties of giant molecular clouds (GMCs), and whether CCCs are actually important for galaxy evolution.

Recent observations have provided substantial evidence supporting the hypothesis that CCCs play a significant role in triggering massive star formation \citep[e.g.][see also Table 1 in \citet{fukui_2021}]{hasegawa_1994,looney_2006,stolte_2008,torii_2011,dewangan_2017_f,nishimura_2018,fujita_2021}. 
In these observations, CCCs are identified through the presence of a bridge feature in position-velocity diagrams, which represents a connection between different velocity components \citep[][]{takahira_2014,haworth_2015_b,haworth_2015_d,torii_2017_a}.
Collision velocities between clouds exhibit a wide distribution, ranging from a few to a few tens of $\kms$, and in one case, even exceeding $100\, \kms$ \citep[see Table 1 in][]{fukui_2021}. 
The observational study conducted by \citet{enokiya_2021_a} has revealed positive correlations between collision velocities and peak column density, as well as between the number of massive stars and peak column densities. These findings suggest that higher column densities are necessary for the formation of massive stars in colliding clouds with higher collision velocities.

Numerical simulations of CCCs have been carried out to study star formation activity and physical properties of gas in colliding clouds \citep[e.g.][]{habe_1992,wu_2017_a,shima_2018,takahira_2018,liow_2020,sakre_2021,sakre_2023,hunter_2023}.
\citet{habe_1992} simulated head-on collisions at supersonic relative velocities between clouds of different sizes and densities. 
Their finding revealed that the shock resulting from the collision compresses the gas at the collision surface, leading to the formation of dense, gravitationally-bound clumps. 
\citet{takahira_2018} also simulated collisions between non-identical clouds with various collision velocities ranging from $5$ to $30\,\kms$.
Although their simulations did not include a prescription of star formation, they found that the fractions of the total dense core mass formed by collisions to the total mass of the colliding clouds decrease with increasing collision speed, suggesting that star formation in colliding clouds is less effective at higher collision speeds.  
This is consistent with the observational implication of \citet{enokiya_2021_a}.

Galaxy simulations have also been used to investigate activities of CCCs on galactic scales \citep[e.g.][]{tasker_2009,tasker_2011,fujimoto_2014_a,fujimoto_2014_b,dobbs_2015,fujimoto_2020, skarbinski_2023}.
In simulations without spiral and/or bar structure, the frequency of CCCs (i.e. how many times a cloud experiences collisions per unit of time) is $30-40\,\gyr^{-1}$ \citep[$\sim1/5 - 1/4$ of the orbital time,][]{tasker_2009,tasker_2011,dobbs_2015}, whilst with imposed spiral arms and/or bar structure the frequency go up to a few hundred $\gyr^{-1}$ \citep[][]{dobbs_2015,fujimoto_2014_a}.
However, in a simulation of a barred spiral galaxy by \citet{fujimoto_2020}, the collision frequency is $10-20\,\gyr^{-1}$, which is much lower than in other simulations.
This may be due to the different galaxy models or the different definitions of clouds, such as density thresholds.

\citet{fujimoto_2014_b} analysed a simulated barred spiral galaxy and found that the collision speed between colliding clouds are more widely distributed in the bar region than in other regions.
In addition, by assuming that the star formation efficiency in colliding clouds varies with collision speed, they reproduced the suppression of star formation in the bar region as reported by the observations \citep[e.g.][]{momose_2010,hirota_2014,yajima_2019,maeda_2020_b}.
Star formation triggered by CCCs, hence, could play an important role in the galaxy evolution.

By solving the evolution equation for the mass function of GMCs, a CCC process in galaxies is also studied.
\citet{kobayashi_2017} found that CCCs are less dominant in determining the GMC mass function, but star formation triggered by CCCs is not included in their equation.
\citet{kobayashi_2018} improved the equation in \citet{kobayashi_2017} by taking the star formation in colliding clouds into account and found that a few tens of per cent of the total star formation rate (SFR) in the Milky Way (MW) and nearby galaxies could be driven by CCCs.

While CCCs have been extensively studied in galaxy simulations, their identification has primarily relied on post-processing analysis. However, if CCCs are indeed capable of promoting star formation, it is reasonable to expect that they would significantly influence the characteristics of subsequent CCCs through stellar feedback. This effect would likely differ from situations where the influence of CCC-induced star formation is not taken into account.

To study the effects of CCC-driven star formation on galaxy evolution, we need to identify CCC events at each timestep, rather than in a post-processing analysis. 
Moreover, the finest spatial resolutions in galaxy simulations are typically on the order of $\pc$, which cannot spatially resolve dense cores whose sizes are $\sim 0.1\,\pc$ \citep[][]{bergin_2007}. 
It is therefore necessary to develop a model that specifically accounts for CCC-triggered star formation and apply it to the CCC events detected on-the-fly during galaxy simulations.

One method for identifying GMCs and CCCs is the Friends-of-Friends (FoF) algorithm.
The FoF is often used to find groups of neighbouring dense gas elements in post-processing analysis of particle-based hydrodynamical galaxy simulations and enables us to investigate the physical properties of GMCs and cloud collisions \citep[e.g.][]{dobbs_2015,pettitt_2018,benincasa_2020}.
Some studies have used an alternative grid-based clump-finding algorithm to identify clouds from snapshots of (particle-based) galaxy simulations \citep[e.g.][]{dobbs_2008,dobbs_2011_b}.
However, \citet{dobbs_2015} pointed out a problem in GMC identification with this approach.
If we compare GMCs identified by the grid-based algorithm at slightly different times in particle-based simulations, their morphology can be quite different.
On the other hand, the shapes of GMCs identified by the FoF do not change significantly over a very short time-scale.
The FoF is, hence, deemed better suited to GMC and CCC identifications than the grid-based method for these kinds of studies in Lagrangian hydrodynamic simulations.

Consequently, in this research, we first develop an algorithm to identify CCCs at each timestep of galaxy simulations.
We then build a model of star formation in colliding clouds from the results of \citet{takahira_2018}, who studied the relationship between the fraction of the core mass in colliding clouds and the collision speed.
We perform hydrodynamic simulations of isolated disc galaxies with the on-the-fly CCC identification algorithm and the CCC-triggered star formation model.
By comparing the simulation results with and without the CCC-triggered star formation model, we investigate how the CCC-triggered star formation impacts star formation in the galaxy and physical properties of GMCs.

This paper is organized as follows.
In Section~\ref{method}, we describe the details of our hydrodynamical simulations of isolated galaxies.
In particular, we introduce how to identify CCCs at each timestep in galaxy simulations and the model of star formation triggered by CCCs.
We present our findings in Section~\ref{result} and give discussions of our results in Section~\ref{discussion}.
Section~\ref{conclusion} concludes our study.

\section{Methods}
\label{method}

\subsection{On-the-fly identification of cloud-cloud collisions}
\label{method:CCC_identification}
In order to apply the star formation model induced by cloud collisions to galaxy simulations, it is necessary to find CCCs at a given time, $t$.
To this end, we first develop an on-the-fly GMC identification algorithm using the FoF at each timestep in galaxy simulations.
We run the FoF algorithm to group neighbouring gas elements with densities above a threshold hydrogen number density, $n_\mathrm{H,min}$, within a linking length, $l_\mathrm{min}$.
If a group contains at least $N_\mathrm{min}$ gas elements, it is identified as a GMC.
We employ $n_{\mathrm{H,min}} = 100\,\cc{}$, $l_{\mathrm{link}} = 10\,\pc{}$, and $N_{\mathrm{min}} = 40$ elements.
We have confirmed that even when these parameters are varied by a factor of two, there are no significant changes in the distribution of physical properties of GMCs, except for the minimum GMC mass and the total number of GMCs.
However, the properties of CCCs depend on the choice of $N_\mathrm{min}$, since the total number of GMCs, and hence the number density of GMCs, strongly depends on $N_\mathrm{min}$.
We define the global properties of each identified GMC.
The position of a GMC, $\bar{\bm{r}}_\mathrm{c}$, is the centre of mass of all the gas elements composing the GMC.
The bulk velocity of a GMC is computed by
\begin{equation}
    \bar{\bm{\varv}}_\mathrm{c} = \langle \bm{\varv} \rangle_M,
\end{equation}
where $\langle X \rangle_M$ denotes a mass-weighted average of a given physical property $X$, $\bm{\varv}$ is the velocity of composite gas elements in a GMC.
The total mass of a GMC, $M_\mathrm{c}$, is the sum of all the gas element masses.
The radius of a GMC, $R_{\mathrm{c}}$, is defined as the radius of a uniform density sphere with the same moment of inertia \citep[][]{guszejnov_2020} using composite gas elements.
We define the 1D velocity dispersion of gas in a GMC, $\sigma_\mathrm{c}$, as
\begin{equation}
    \sigma_\mathrm{c}^2 = \frac{1}{3} \left( \langle (\bm{\varv} - \bar{\bm{\varv}}_\mathrm{c})^2 \rangle_M \right).
\end{equation}
The sound speed of a GMC, $c_{\mathrm{s,c}}$, is given as
\begin{equation}
    c_{\mathrm{s,c}} = \sqrt{\gamma(\gamma-1)\langle u \rangle_M},
\end{equation}
where $u$ is the specific internal energy of composite gas elements in a GMC and $\gamma = 5/3$ is the adiabatic index.
We also define the virial parameter, $\alpha_\mathrm{vir}$, of a GMC as
\begin{equation}
    \alpha_\mathrm{vir} = \frac{5(\sigma_\mathrm{c}^2 + c_{\mathrm{s,c}}^2)R_{\mathrm{c}}}{GM_{\mathrm{c}}}, \label{eq:virialCloud}
\end{equation}
where $G$ is the gravitational constant \citep[][]{bertoldi_1992}.
The minimum mass of the identified GMCs is $\sim 10^4\,\Msun$ with the original mass of a gas element, $250 \, \Msun$ (see Sec.~\ref{method:numerical_simulations}), and $N_\mathrm{min} = 40$.
This is a typical mass of the smallest GMCs which hosts star formation \citep[e.g.][]{williams_2000, inoue_2013, kobayashi_2017} and is sufficiently low to investigate the star formation in the colliding GMCs (see Sec.~\ref{result:sf:sfr} and \ref{result:gmc:ccc_event}).

To identify CCCs on-the-fly, we need not only the information of GMCs at time $t$, but also the information at the previous timestep $t - \Delta t$, where $\Delta t$ represents the timestep\footnote{The timestep $\Delta t$ in this paper indicates the system timestep as the minimum of the hierarchical timesteps of all particles. The timestep in our simulations is $\sim 100-1000\,\yr$.}.
To find GMCs at $t - \Delta t$, we store the physical properties of gas elements such as positions, velocities, masses, densities, and internal energies at $t - \Delta t$ for all gas elements\footnote{We store the drifted values at $t - \Delta t$ for inactive gas elements.}.
Once we have identified the GMCs at time $t$, we apply the FoF algorithm to the positions of the gas elements at $t - \Delta t$ in order to find the GMCs at time $t - \Delta t$.
We, hence, have to run the GMC identification algorithm twice at every timestep.
In this paper, the physical properties of GMCs found at $t - \Delta t$ are denoted with a superscript $\mathrm{p}$ (e.g. $\bar{\bm{r}}_\mathrm{c}^\mathrm{p}, \bar{\bm{\varv}}_\mathrm{c}^\mathrm{p}, M_\mathrm{c}^\mathrm{p}$ etc.).
By comparing the identified GMCs at $t$ and $t - \Delta t$,  we can determine how GMCs have evolved, i.e. whether the GMCs collide or not.
To determine CCCs, we impose two criteria as follows.

\begin{figure*}
    \begin{center}
        \includegraphics[width=2\columnwidth]{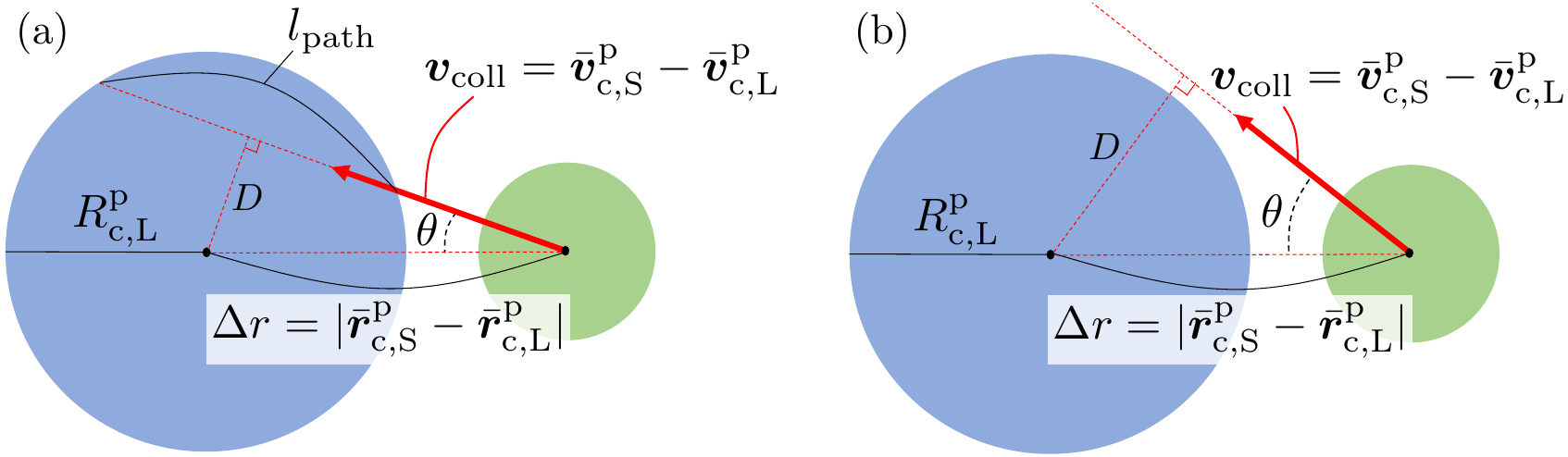}
        \caption{
                Schematic illustrations of how to determine a colliding GMC which would trigger star formation in our galaxy simulations.
                A collision is determined using the physical properties of merging GMCs at $t-\Delta t$, assuming that the GMCs are spherical and move linearly.
                The subscripts $\mathrm{L}$ and $\mathrm{S}$ denote GMCs with larger and smaller GMC radii $R_\mathrm{c}^\mathrm{p}$, respectively.
                To identify a collision, we calculate the following values: $\Delta r$ is the distance between the centres of mass of GMCs, $\bm{\varv}_\mathrm{coll}$ is the collision velocity, $\theta$ is the collision angle, and $D$ is the distance from the centre of mass of the larger GMC to the line along the collision velocity.
                (a) The case identified as a colliding GMC for $\Delta r > R_\mathrm{c,L}^\mathrm{p}$.
                    The centre of mass of the smaller GMC penetrates the larger one and the composite gas is expected to be compressed sufficiently to trigger star formation.
                    How far the centre of mass of the smaller GMC passes through the larger one, $l_\mathrm{path}$, is calculated by using $\Delta r, \theta, R_\mathrm{c,L}^\mathrm{p}$, and $D$.
                    We then define the collision time-scale $\Delta t_\mathrm{coll} = l_\mathrm{path}/\varv_\mathrm{coll}$, during which we apply the model of star formation triggered by CCCs.
                (b) The case not identified as a colliding GMC.
                    Although these GMCs could also become a merging GMC at $t$ by the first criterion for determining a CCC, they would just scratch each other and we assume that sufficient gas compression does not occur.
                }
        \label{fig:headon_collision}
    \end{center}
\end{figure*}

\begin{enumerate}
    \item \label{ccc_cond1}
    We determine whether a GMC qualifies as a merging cloud at a given time, $t$, using the methodology employed by \citet{dobbs_2015} in their post-processing analysis. 
    When examining each GMC detected at time $t$, we first define $f_i$ as the fraction of that GMC that is derived from a cloud identified at a previous timestep, specifically $t - \Delta t$. 
    We then arrange these $f_i$ values in descending order, ensuring that $f_1 \geq f_2 \geq f_3 \dots$. 
    If both $f_1$ and $f_2$ have non-zero values (indicating the presence of at least two progenitors), we classify the GMC as a candidate of a \textit{merging} cloud.\footnote{Figure~3 of \citet{dobbs_2015} may help to understand this procedure.}

    To prevent the misclassification of a GMC as a merging cloud when the contributions from GMCs at $t - \Delta t$ are minimal, we have implemented additional criteria for identifying GMCs as merging clouds:
    \begin{equation} 
        f_1 M_\mathrm{c} \geq f_\mathrm{th} M_\mathrm{c,1}^\mathrm{p} 
    \end{equation} 
    and 
    \begin{equation} 
        f_2 M_\mathrm{c} \geq f_\mathrm{th} M_\mathrm{c,2}^\mathrm{p}, 
    \end{equation} 
    where $f_\mathrm{th}=0.5$ is the threshold fraction and $M_\mathrm{c,1}^\mathrm{p}$ and $M_\mathrm{c,2}^\mathrm{p}$ denote the progenitor GMC masses at $t-\Delta t$ that contribute the mass fractions $f_1$ and $f_2$ to the GMC at time $t$, respectively. 
    These criteria ensure that at least half of the progenitor mass is contained within the cloud identified as a merging cloud.
    \item \label{ccc_cond2}
        When two GMCs collide in a grazing manner, the gas within the merging cloud is unlikely to be compressed sufficiently to induce intense star formation.
        Since one of our goals in this study is to find GMC mergers that would lead to star formation in galaxy simulations, we now impose an additional criterion for finding such mergers.
        Fig.~\ref{fig:headon_collision} shows the schematic illustrations of how to determine a GMC merger that would lead to star formation. 
        We assume that GMCs just before merging (i.e. at $t-\Delta t$) are spherical and denote a GMC with larger and smaller $R_\mathrm{c}^\mathrm{p}$ as the subscripts $\mathrm{L}$ and $\mathrm{S}$, respectively.
        We calculate the relative position of GMCs at $t-\Delta t$ as
        \begin{equation}
            \Delta\bm{r} =\bar{\bm{r}}_\mathrm{c,S}^\mathrm{p} - \bar{\bm{r}}_\mathrm{c,L}^\mathrm{p}.
        \end{equation}
        The cloud collision velocity, $\bm{\varv}_\mathrm{coll}$, is calculated from the velocities of the GMCs at $t-\Delta t$, $\bar{\bm{\varv}}_\mathrm{c,L}^\mathrm{p}$ and $\bar{\bm{\varv}}_\mathrm{c,S}^\mathrm{p}$:
        \begin{equation}
            \bm{\varv}_\mathrm{coll} =\bar{\bm{\varv}}_\mathrm{c,S}^\mathrm{p} - \bar{\bm{\varv}}_\mathrm{c,L}^\mathrm{p}.
        \end{equation}
        We are now able to define the collision angle, $\theta$, as
        \begin{equation}
            \cos(\pi-\theta) = \frac{\Delta\bm{r} \cdot \bm{\varv}_\mathrm{coll}}{\Delta r\, \varv_\mathrm{coll}},
        \end{equation}
            where $\Delta r = |\Delta\bm{r}|$ and $\varv_\mathrm{coll} = |\bm{\varv}_\mathrm{coll}|$.
        We, here, define the distance from the centre of mass of the larger GMC to the line along the collision velocity, $D$, as
        \begin{equation}
            D = \Delta r \sin\theta.
        \end{equation}
        If $D < R_\mathrm{c,L}$, this merger is expected to compress the composite gas and form stars since the centre of mass of the smaller GMC can penetrate the larger one assuming linear motion (see Fig.~\ref{fig:headon_collision}a).
        Otherwise, the GMCs would just scratch each other and gas would not be compressed sufficiently to trigger star formation (see Fig.~\ref{fig:headon_collision}b).
        Although we discuss above how to identify GMC mergers that trigger star formation as if $\Delta r > R_\mathrm{c,L}^\mathrm{p}$, where the GMCs appear to be separated at $t-\Delta t$, $\Delta r < R_\mathrm{c,L}^\mathrm{p}$ is possible in this estimate due to the assumption that both GMCs are spherical.
        In this case, we employ the criterion, $\theta < \pi/2$, to reflect that the GMCs are approaching each other, instead of $D < R_\mathrm{c,L}$.
\end{enumerate}

GMCs that fulfil the above two criteria are identified as colliding GMCs in which star formation would be promoted.
To determine how long a collision event lasts, we define $l_\mathrm{path}$, which denotes how far the centre of mass of the smaller GMC passes through the larger one assuming linear motion.
If $\Delta r > R_\mathrm{c,L}^\mathrm{p}$, $l_\mathrm{path}$ is calculated by
\begin{equation}
    l_\mathrm{path} = 2\sqrt{\left(R_\mathrm{c,L}^\mathrm{p}\right)^2 - D^2},
\end{equation}
while if $\Delta r < R_\mathrm{c,L}^\mathrm{p}$, $l_\mathrm{path}$ is computed as
\begin{equation}
    l_\mathrm{path} = \Delta r \cos\theta + \sqrt{\left(R_\mathrm{c,L}^\mathrm{p}\right)^2 - D^2}.
\end{equation}
During $\Delta t_\mathrm{coll} = l_\mathrm{path}/\varv_\mathrm{coll}$ or until the colliding GMC is destroyed, we apply the star formation model by CCCs to the gas elements making up the colliding GMCs.
We output the information on the physical properties of colliding GMCs at $t$ and GMCs just before collisions at $t-\Delta t$ when collisions are detected.
This enables us to analyse the actual collision properties and the differences between the properties of CCCs identified by the on-the-fly algorithm and those found by the post-processing analysis.

\subsection{Modelling star formation triggered by cloud-cloud collisions}
\label{method:model_SF_CCC}
Since dense cores and individual stars in colliding GMCs cannot be spatially resolved in our galaxy simulation, we need to model star formation triggered by CCCs.
In galaxy-scale simulations, star formation is often parameterized by
\begin{equation}
    \dot{\rho}_\star = \epsilon_\mathrm{ff,SF}\frac{\rho}{t_\mathrm{ff}},
\end{equation}
    for star-forming gas elements, where $\dot{\rho}_\star$ is the SFR density, $\rho$ is the local gas density, $t_\mathrm{ff}=\sqrt{3\pi/32G\rho}$ is the free-fall time, and $\epsilon_\mathrm{ff,SF}$ is the star formation efficiency per free-fall time parameter.
In many previous studies, a constant $\epsilon_\mathrm{ff,SF}$ is adopted in galaxy simulations \citep[e.g.][]{agertz_2013,okamoto_2017,hopkins_2018_b}, while some employed a variable $\epsilon_\mathrm{ff,SF}$ in isolated disk galaxy and cosmological simulations \citep[e.g.][]{semenov_2016,li_2018}.
In this research, we develop a simple model of a variable $\epsilon_\mathrm{ff,SF}$ that reflects the triggered star formation in colliding GMCs.

\citet{takahira_2018} simulated the colliding clouds with a relatively wide range of collision velocities from $5$ to $30\,\kms$.
They found that the fractions of the total dense core mass formed via collisions to the total cloud mass at the converging points tend to decrease as the collision velocities increase (see Table~2 in \citet{takahira_2018}).
However, their simulations did not include a prescription of star formation.
On the other hand, \citet{shima_2018} also simulated colliding clouds at $10$ and $20\,\kms$ collision speeds, including the formation of sink particles to represent star formation.
They found that in the $10\,\kms$ case, the fraction of the total mass of the sink particles to the total mass of the cloud is $\sim 10$ per cent at $6\,\myr$ after the collision, which is approximately equal to the free-fall time of typical GMCs.
This fraction is about half of the core mass fraction in \citet{takahira_2018}, although the time-scale is different.
We, hence, assume that $50$ per cent of the dense core mass is converted into stars per free-fall time and make a model of $\epsilon_\mathrm{ff,SF}$ by fitting the result of \citet{takahira_2018} with an exponential function of the cloud collision speed $\varv_\mathrm{coll}$ using the least squares method.
Then we get\footnote{We obtain Eq.~(\ref{eq:epsilon_coll}) by fitting the results of \citet{takahira_2018} with $\ln(\epsilon_\mathrm{ff,SF})$, which is a linear function of $\varv_\mathrm{coll}$. The correlation coefficient for this best fit is $-0.92$.}
\begin{equation}
    \epsilon_\mathrm{ff,SF} = 0.32 \exp\left(-0.093\frac{\varv_\mathrm{coll}}{1\,\kms}\right).
    \label{eq:epsilon_coll}
\end{equation}
To avoid star formation in colliding GMCs whose collision speed is not high enough to sufficiently compress the composite gas, we impose the lower limit for $\varv_\mathrm{coll}$, $\varv_\mathrm{coll} > c_\mathrm{s,c}$, where $c_\mathrm{s,c}$ is the sound speed of a GMC at $t$.
The sound speed $c_\mathrm{s,c}$ is typically $\lesssim1\,\kms$, which is usually slower than the collision speed (see Section~\ref{result:gmc:gmc_properties}).
While the simulations of CCCs by \citet{takahira_2018} covered the collision speeds between $5$ and $30\,\kms$, we apply this $\epsilon_\mathrm{ff,SF}$ to gas elements in colliding GMCs even when the collision speeds are beyond the range.
Although other factors, such as density structure, morphology, turbulence of the GMCs etc., may affect star formation in colliding GMCs, we here take into account only $\varv_\mathrm{coll}$ for simplicity.
For dense gas elements which are not members of colliding GMCs, we take $\epsilon_\mathrm{ff,SF}=0.01$ since this value is comparable to previous galaxy simulations \citep[][]{okamoto_2017,fujimoto_2019}, although much larger value of $\epsilon_\mathrm{ff,SF}=1$ has been also used \citep[e.g.][]{hopkins_2018_b}.
We note that $\epsilon_\mathrm{ff,SF}$ is different from the star formation efficiency per free-fall, $\epsilon_\mathrm{ff}$, which is estimated for star-forming GMCs \citep[see also][]{grisdale_2019}.
Our model is designed to account for the promoted star formation in colliding GMCs depending on collision speeds in galaxy simulations.

In our simulations, gas elements with temperature $T<100\,\K$ and hydrogen number density $n_\mathrm{H}>100\,\cc$ are stochastically converted into star particles using $\epsilon_\mathrm{ff,SF}$, following \citet{okamoto_2014}.
Since our simulations do not have high enough resolution to resolve individual stars, we assume that each star particle follows a simple stellar population of the Chabrier initial mass function \citep[][]{chabrier_2003} in the range from $0.1$ to $100\,\Msun$.

\subsection{Stellar feedback}
\label{method:stellar_feedback}
After star particles are born, they chemically and dynamically affect surrounding gas elements through stellar feedback, including stellar winds, core-collapse (CC) and Type Ia supernovae (SNe), stellar mass loss from AGB stars,  and radiation from massive stars. 
We adopt the metallicity-dependent stellar lifetime by \citet{portinari_1998} to compute timed release of mass, metals, and energy from a star particle.

We compute the energy released as stellar winds by using tables generated by STARBURST99 \citep{leitherer_1999_starburst99}. 
Note that here we ignore the mass-loss of massive stars as it is included in the ejecta mass of the core-collapse SNe (CCSNe) in the yield tables compiled by \citet{nomoto_2013}. 
We also tabulate the number of ionising photons and the luminosity of an SSP as a function of its age and metallicity by using a stellar population synthesis code P\'{E}GASE.2 \citep{fioc_1997_pegase} to implement radiative feedback. 
For CCSNe, we compute the released mass, metals, and energy by using the yield tables complied by \cite{nomoto_2013}. 
In \citet{nomoto_2013}, stars from 13 to 40~$\Msun$ (from 13 to 300~$\Msun$ for the zero metal stars) are assumed to explode as CCSNe.
We lower the minimum mass to a canonical value of $8~\Msun$ by extrapolating the tables.
As a result, the number of CCSNe from a star particle is twice as many as the original assumption. 
For Type Ia SNe, we assume a power-law form of the delay-time distribution function taken from \citet{maoz_2012} for a star particle older than $4\times 10^7$ yr. 
We employ the metallicity-dependent yield as the N100 model of \citet{seitenzahl_2013}. 
For the AGB yield, we combine the yield tables by \cite{campbell_2008}, \cite{karakas_2010}, \citet{gil-pons_2013}, and \citet{doherty_2014_a}, as in \citet{saitoh_2017}. 

For stellar winds and SNe feedback, we first compute a vector weight, $\bm{\varw}_{\alpha j}$, between a star particle, $\alpha$, and a gas element, $j$, following the method described in \citet{hopkins_2018_a}.
The sum of $\bm{\varw}_{\alpha j}$ and $|\bm{\varw}_{\alpha j}|$ over $j$ are taken to be $0$ and $1$, respectively.
When the feedback energy, $E_\alpha$, is released from a star particle, $\alpha$, the neighbours receive the energy proportional to $|\bm{\varw}_{\alpha j}|$ as thermal energy.
The energy added to a gas element, $j$, from the star particle, $\alpha$, is given by
\begin{equation}
    \Delta E_{\alpha j} = |\bm{\varw}_{\alpha j}| E_\alpha.
\end{equation}
The mass added to a gas element is also calculated in a similar way:
\begin{equation}
    \Delta m_{\alpha j} = |\bm{\varw}_{\alpha j}| M^\mathrm{ej}_\alpha,
\end{equation}
    where $M^\mathrm{ej}_\alpha$ is the ejecta mass.
As above, we employ pure thermal feedback to model the SNe feedback. 
Although the cooling radius is not fully resolved with the resolution of $m_\mathrm{gas} = 250 \, \Msun$\footnote{The kernel size of a gas cell at the star formation threshold density ($n_\mathrm{H} = 100\,\mathrm{cm}^{-3}$) is $\sim 8$ pc, while the cooling radius at this density is $\sim 4$ pc.}, pure thermal feedback has a noticeable effect (see e.g. Fig.~\ref{fig:gas_map}). 
We, however, note that our simulations may underestimate the feedback effect.

A young stellar population, characterized by a young star particle, emits ionising radiation. 
This radiation creates an \HII{} region and may affect the star formation. 
Gas elements inside $\HII$ regions are stochastically heated to $10^4\,\K$ following \citet{marinacci_2019}, instead of solving the radiation transfer equation with resolved \HII{} regions.
We forbid heated gas to cool below $10^4\,\K$ until the end of the timestep of the young stellar particle or the occurrence of the first supernova of the massive star, whichever comes first.
The radiation from a young stellar population also has an additional impact on the surrounding gas \citep{hopkins_2011, agertz_2013, okamoto_2014, ishiki_2017}. 
In our simulations, we inject momentum around a young star particle of the form
\begin{equation}
    p^\mathrm{rad}_\alpha = \frac{L_\alpha}{c} (1 + \tau_{\mathrm{IR}, \alpha})\Delta t_\alpha, 
\end{equation}
where $p^\mathrm{rad}_\alpha$ is the total radial momentum injected by the star 
particle over the timestep $\Delta t_\alpha$, 
$L_\alpha$ is the luminosity of the star particle, $c$ is the speed of light 
and $\tau_\mathrm{IR}$ is the optical depth of the surrounding gas to infrared radiation. 
The optical depth is calculated as $\tau_{\mathrm{IR}, \alpha} = \kappa_\mathrm{IR} \Sigma_{\mathrm{gas}, \alpha}$ where 
$\kappa_\mathrm{IR}$ and $\Sigma_{\mathrm{gas}, \alpha}$ are the opacity in the infrared wavelength and 
the gas column density respectively. 
We employ $\kappa_\mathrm{IR} = 10 (Z/0.02)~\mathrm{cm}\, \mathrm{g}^{-1}$ \citep{hopkins_2018_b} and we estimate the gas column density locally by the Soblev approximation.

\subsection{Numerical simulations}
\label{method:numerical_simulations}
Simulations are performed using the GIZMO code \citep[][]{hopkins_2015_gizmo}.
The self-gravity is solved by the solver inherited from GADGET-3 \citep[][]{springel_2005_gadget2} and the hydrodynamics is solved by meshless finite mass (MFM) method.
The softening length for star particles is set to $10\,\pc$, while that for gas elements is fully adaptive with a minimum value of $1\,\pc$ \citep[][]{price_2007}.

We distribute the gas elements in a static logarithmic potential to represent a MW-like flat rotation curve.
The form of the potential is written as
\begin{equation}
    \Phi(R_\mathrm{gal},z_\mathrm{gal}) = \frac{1}{2} v_0^2 \ln\left(R_\mathrm{gal}^2 + R_0^2 + \frac{z_\mathrm{gal}^2}{q_\Phi^2}\right),
\end{equation}
    where $R_\mathrm{gal}$ and $z_\mathrm{gal}$ represent the galactic radii and the vertical height, respectively, and the parameters are set to $v_0 = 220 \, {\mathrm{km\,s^{-1}}}$, $R_0 = 1 \, \kpc{}$ and $q_\Phi = 0.35$ \citep[][]{binney_galactic_dynamics2}.
We initially distribute the gas elements so that the gas surface density follows the Toomre $Q$ values \citep{toomre_1964} as a function of the galactic radii as follows:
\begin{equation}
    Q = 
    \left\{
        \begin{array}{ll}
            20 & (R_\mathrm{gal} < 2\,\kpc{}) \\
            1 & (2\,\kpc{} \leq R_\mathrm{gal} < 13\,\kpc{}) \\
            20 & (13\,\kpc{} \leq R_\mathrm{gal} < 14\,\kpc{})
        \end{array}
        \right.
        ,
\end{equation}
with the vertical gas distribution of ${\mathrm{sech}}^2(z_\mathrm{gal}/z_h)$ where $z_h = 0.4 \,\kpc$.
This initial setup is similar to the previous simulations of isolated galaxies \citep[e.g.][]{tasker_2009, fujimoto_2019}. 
The initial total gas mass is $8.6\times10^9\,\Msun$ and the number of the gas elements is $34.4$ million, giving an initial gas mass resolution of $250\,\Msun$.
The gas initially has a temperature\footnote{The temperature is determined so as to give the $Q$ values in combination with the total gas mass and the rotation curve and corresponds to the sound speed of $\sim 5.4\,\kms$.} of $2540\,\K$ and solar metallicity, $Z_{\sun} = 0.013$ \citep[][]{asplund_2009}.

We use pre-computed tables to calculate radiative cooling and heating following \citet{hopkins_2023}, which is the updated version of \citet{hopkins_2018_b}.
We include the fine-structure and molecular cooling down to $10\,\K$, the cooling rates dependent on the metal species lines \citep[][]{wiersma_2009}, and the dust-gas collisional heating and cooling \citep[][]{meijerink_2005} as in \citet{hopkins_2018_b}.
The molecular hydrogen fractions are estimated according to \citet{krumholz_2009}.
We take into account the self-shielding from the UV background by using the fitting function of \citet{hopkins_2023} \citep[see also][]{rahmati_2013}.
We also include photoelectric heating with a constant rate of $8.5 \times 10^{-26} \mathrm{erg \, s^{-1}}$ per hydrogen atom, as in \citet{fujimoto_2019}.
A pressure floor is added to prevent gas elements from collapsing to very high densities beyond the threshold density for star formation.
The minimum pressure is calculated as $P_\mathrm{f}=P_0(n_\mathrm{H}/n_0)^\gamma$, where $P_0$ is the pressure of the gas with temperature $T=100\,\K$ and the hydrogen number density $n_\mathrm{H}=n_0=100 \, \cc$, and $\gamma=5/3$ is the adiabatic index.
Note that this pressure floor is only applied to the gas with $n_\mathrm{H} > n_0$ to prevent star forming gas from collapsing in dynamical time. Furthermore, it should be emphasized that this pressure floor is unlikely to have a significant effect on the formation of giant GMCs, since it applies exclusively to high-density gas.
In addition, we have confirmed that the internal velocity dispersion of GMCs is typically much larger than the sound speed calculated from the imposed pressure. As a result, we expect that the properties of GMCs, such as their sizes and virial parameters, are not strongly affected by the pressure floor.
We employ the explicit metal diffusion between gas elements following \citet{colbrook_2017} and \citet{hopkins_2018_b}.

We first run only the simulation with $\epsilon_\mathrm{ff,SF}=0.01$ for $300\,\myr$ to evolve the galactic disc.
We then turn on the on-the-fly CCC identification algorithm and the model of star formation triggered by CCCs introduced in Sec.~\ref{method:CCC_identification} and \ref{method:model_SF_CCC} until $t=500\,\myr$.
This simulation is denoted as "\textit{Coll}".
We also run the simulation with a constant $\epsilon_\mathrm{ff,SF}$ of $0.01$ even if gas elements are parts of colliding GMCs, which is denoted as "\textit{Const}".
In the Const simulation, we turn on the on-the-fly CCC identification algorithm as well, in order to obtain information about CCCs in the galaxy without the star formation model of CCCs.
By comparing the results of the Coll and Const simulations, we investigate how star formation triggered by CCCs affects galaxy evolution and GMC properties.
We analyse the data between $t=350\,\myr$ and $500\,\myr$ to avoid starbursts just after turning on the star formation model of CCCs in the Coll simulation.
Our analysis is restricted to the main disc region of $2\,\kpc < R_\mathrm{gal} < 13\,\kpc$ to prevent the effect of the artificially lowered gas surface density in the initial condition.
However, its impact on our results is small.

\section{Results}
\label{result}

\subsection{Star formation}
\label{result:sf}

\begin{figure*}
    \begin{center}
        \includegraphics[width=2\columnwidth]{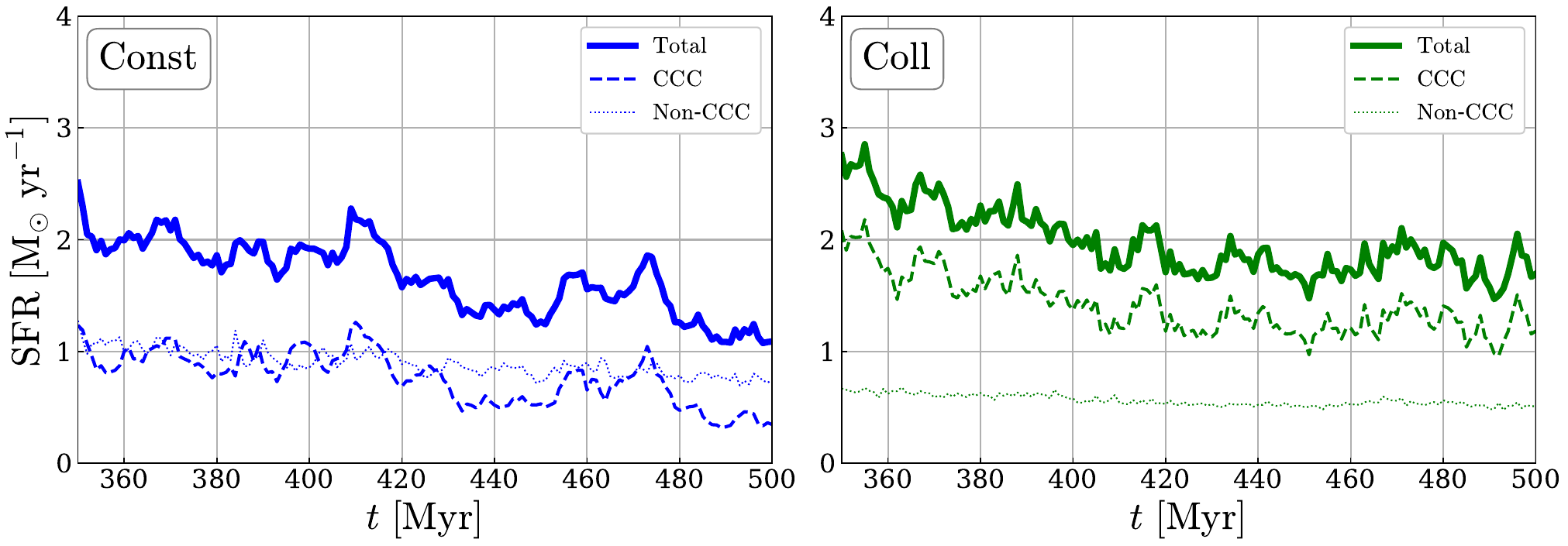}
        \caption{
            SFRs from $350-500\,\myr$ in the Const (left) and Coll (right) simulations.
            The SFRs are calculated from young star particles with ages $<1\,\myr$.
            The thick solid lines are the SFRs calculated from all young star particles (Total).
            The dashed lines are from young star particles born in colliding clouds (CCC).
            The SFRs from young star particles born outside the colliding clouds are indicated by the thin dotted lines (Non-CCC).
        }
        \label{fig:sfh_cmp_coll}
    \end{center}
\end{figure*}

\subsubsection{Star formation rates}
\label{result:sf:sfr}

We first show the effect of star formation triggered by CCCs on the SFRs of the simulated galaxies.
Fig.~\ref{fig:sfh_cmp_coll} shows the time evolution of the SFRs in the Const and Coll simulations from $t=350-500\,\myr$.
We denote the three types of SFRs as Total, CCC, and Non-CCC.
The first one ($\mathrm{SFR_{Total}}$) is obtained from all young star particles, and the second ($\mathrm{SFR_{CCC}}$) is estimated using young stars born in colliding GMCs.
The last one ($\mathrm{SFR_{Non-CCC}}$) represents star formation occurring outside the colliding GMCs (i.e. \mbox{$\mathrm{SFR_{Non-CCC}} = \mathrm{SFR_{Total}} - \mathrm{SFR_{CCC}}$}).
The SFRs are calculated using star particles younger than $1\,\myr$.
We find that the Coll simulation produces slightly higher Total SFRs than the Const simulation, suggesting that CCCs promote star formation activity on galactic scales.
The Total SFRs in both simulations gradually decrease by $\sim 1.0-1.3\,\sfrunit$ over the analysis period of $150\,\myr$.
On average, these values are $\sim 1.7\,\sfrunit$ in Const and $\sim 2.0\,\sfrunit$ in Coll.
This means that approximately $\sim20$ per cent more stars are born when CCC-triggered star formation is taken into account.
However, this change does not suggest that $20$ per cent of stars are born via CCCs.

Focusing on the CCC SFRs, the values in Coll appear to be higher than those in Const.
The CCC SFRs in the Const and Coll simulations are on average $\sim 0.77\,\sfrunit$ and $\sim 1.4\,\sfrunit$, respectively.
Although the CCC SFR is $\sim 0.6\,\sfrunit$ higher in Coll than in Const, the difference in the Total SFR between the simulations is smaller than in the CCC SFRs.
This result suggests that star formation is self-regulated by stellar feedback.
Previous studies have shown that stellar feedback from young stars destroys surrounding dense gas and suppresses star formation \citep[e.g.][]{hopkins_2011,agertz_2013,colling_2018,chevance_2022}.
The feedback regulation of star formation probably weakens the dependence of Total SFRs on the star formation model of CCCs.

The fractions of the CCC SFRs to the Total SFRs are $\sim 50$ per cent in Const and $\sim 70$ per cent in Coll.
The fraction in the Const simulation is consistent with the previous study of \citet{kobayashi_2018}, who solved a modelled equation of GMC evolution including star formation driven by CCCs and found that CCCs could cover $20-50$ per cent of galactic SFRs.
On the other hand, the fraction in the Coll simulation is $\sim 20-50$ per cent higher than the estimate from the semi-analytical evolutionary GMC description.
We note that our simulations and \citet{kobayashi_2018} use different methodologies, such as numerical methods, parameters, star formation models, and so on. 
These differences could affect the fractions of CCC SFRs in various scenarios. 
Since exploring such differences is beyond the scope of this work, we here just mention that the Coll simulation produces a higher fraction of the CCC SFRs by a few tens of per cent than the previous study.

Interestingly, the fluctuations in the Total SFRs in both simulations 
closely mirror those in the CCC SFRs.
Throughout the analysis period, the Non-CCC SFRs exhibit minimal immediate fluctuations.
This result suggests that CCCs play an important role in determining large instantaneous changes in the galactic total SFRs.
Regardless of whether the CCC triggering star formation model is employed, gas is compressed through the CCC, leading to a sudden SFR increase.

\begin{figure}
    \begin{center}
        \includegraphics[width=\columnwidth]{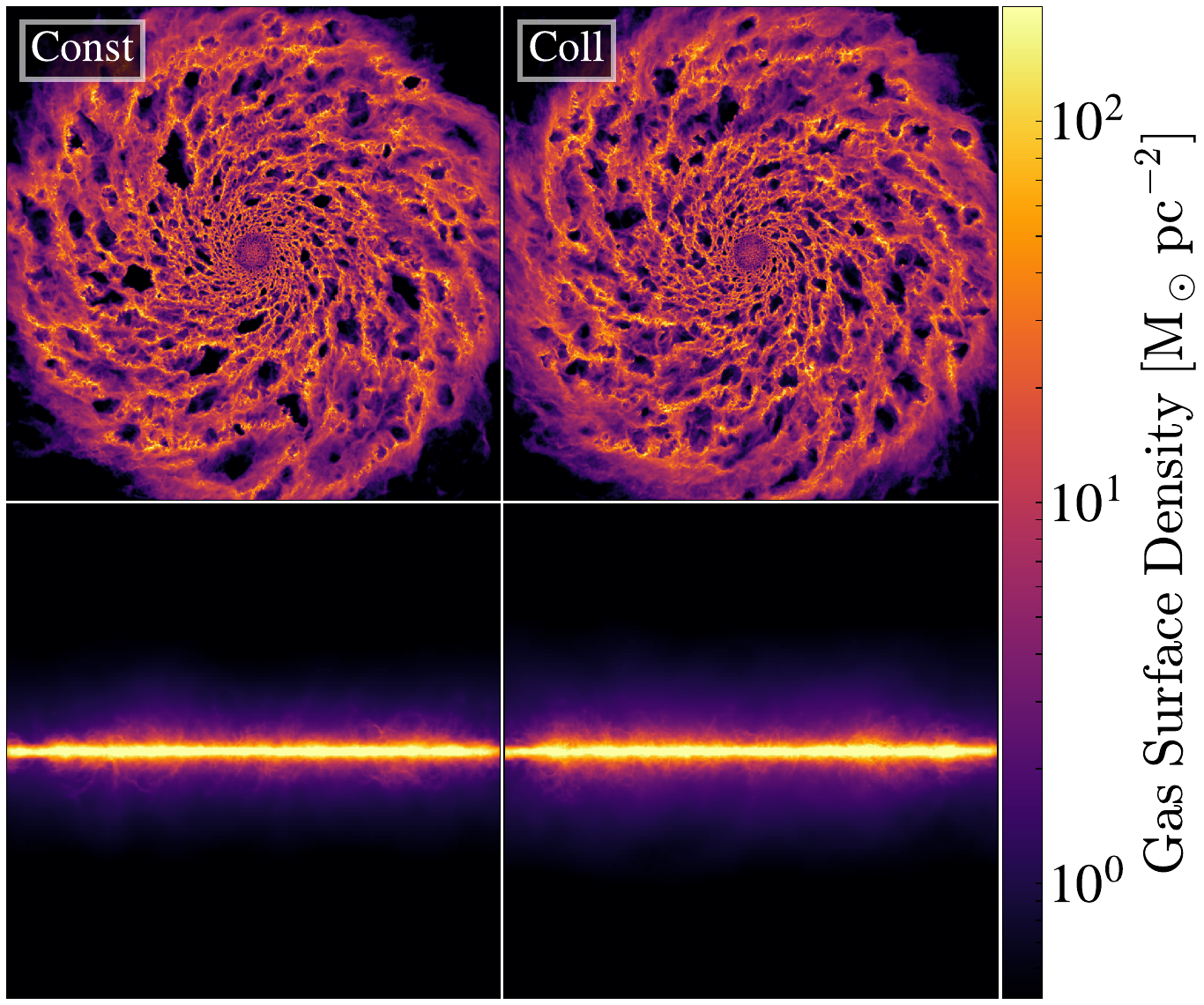}
        \caption{
            Face-on (top) and edge-on (bottom) views of the gas surface density maps of the galaxies at $t=500\,\myr$ for Const (left) and Coll (right).
            Each panel is $26\,\kpc$ across.
        }
        \label{fig:gas_map}
    \end{center}
\end{figure}

In Fig.~\ref{fig:gas_map}, we show the face-on and edge-on views of the gas surface density of the simulated galaxies at $t = 500\,\myr$.
They are visually very similar, including 
spiral arms, sizes of cavities ($\sim \kpc$), and gas scale height.
This similarity implies that the CCC-triggered star formation model has little impact on the gas distribution at galactic scales. In the Coll simulation, the Total SFR surpasses that of the Const simulation by a mere 20 per cent, indicating that enhancement of the stellar feedback appears to be relatively insubstantial.

\begin{figure}
    \begin{center}
        \includegraphics[width=\columnwidth]{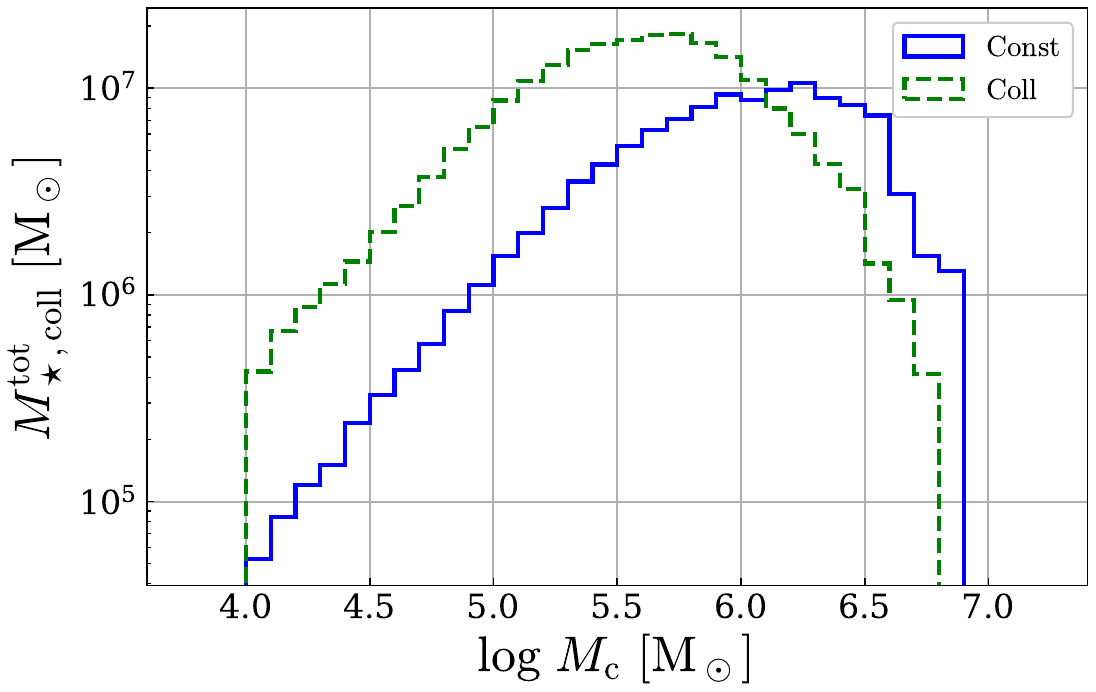}
        \caption{
            Total masses of stars born between $350-500\,\myr$ in colliding GMCs as a function of the mass of the host colliding GMCs.
        }
        \label{fig:Mstar_vs_Mcloud_collision}
    \end{center}
\end{figure}

In Fig.~\ref{fig:Mstar_vs_Mcloud_collision}, we show the total masses of stars born in colliding GMCs, $M_\mathrm{\star,coll}^\mathrm{tot}$, as a function of the masses of the colliding GMCs where the stars are born, $M_\mathrm{c}$.
In the range of $M_\mathrm{c} \lesssim 10^6\,\Msun$, the colliding GMCs in Coll form several times more stars than in Const.
The distributions have the peaks of $M_\mathrm{\star,coll}^\mathrm{tot} \sim 2\times10^7\,\Msun$ at $M_\mathrm{c} \sim 10^{5.7}\,\Msun$ for the Coll simulation and $M_\mathrm{\star,coll}^\mathrm{tot} \sim 10^7\,\Msun$ at $M_\mathrm{c} \sim 10^{6.2}\,\Msun$ for the Const simulation.
In the range above the peak in each simulation, $M_\mathrm{\star,coll}^\mathrm{tot}$ steeply declines due to the smaller number of massive GMCs (see Sec.~\ref{result:gmc:gmc_properties}).
It is important to note that the Const simulation uses a fixed value of $\epsilon_\mathrm{ff,SF} = 0.01$, but the choice of this parameter can have a discernible effect on the star formation activity within colliding GMCs. 
\citet{li_2020} found that using a higher value of $\epsilon_\mathrm{ff,SF}$ reduces the likelihood of producing larger GMCs. 
Consequently, an increase in $\epsilon_\mathrm{ff,SF}$ may result in a reduced number of stars being born within the more massive colliding GMCs, potentially leading to a decrease in the fraction of CCC-induced SFRs relative to total SFRs.

\subsubsection{Kennicutt-Schmidt law}
\label{result:sf:ks_law}

\begin{figure*}
    \begin{center}
        \includegraphics[width=2\columnwidth]{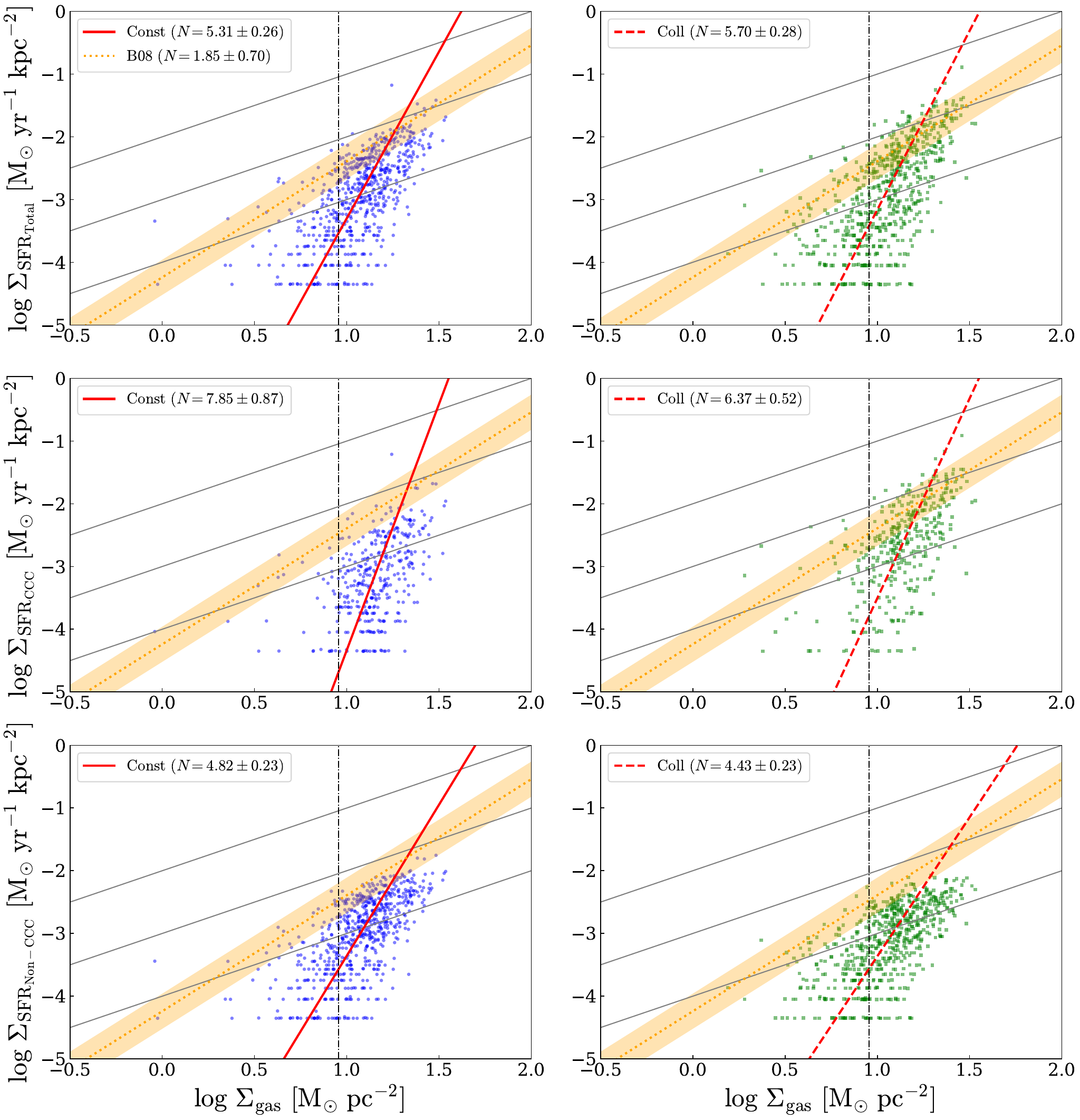}
        \caption{
            Dependence of star formation in colliding GMCs on the Kennicutt-Schmidt relations.
            The left side and right side panels show the Const and Coll results at $t=500\,\myr$, respectively. 
            The area-averaged SFR, $\Sigma_\mathrm{SFR}$, and the gas surface density, $\Sigma_\mathrm{gas}$, are calculated in $750\,\pc \times 750\,\pc$ pixels.
            The SFRs in these plots are calculated using star particles younger than $10\,\myr$.
            We estimate SFRs in the top, middle, and bottom panels with all young stars, young stars born in colliding GMCs, and those born outside the colliding GMCs, respectively. 
            We fit the results to $\Sigma_\mathrm{SFR} \propto \Sigma_\mathrm{gas}^N$ and $N$ for each analysis is listed in the legends.
            For comparison, we show an observational result \citep[][B08]{bigiel_2008}, depicted as orange dotted lines and accompanied by orange shading denoting the $1\sigma$ error. 
            The vertical black solid lines at $\Sigma_\mathrm{gas}=9 \,\Msun$ indicate the gas surface density threshold that marks the transition between \HI{} and $\mathrm{H_2}$ dominance according to \citet[][B08]{bigiel_2008}.
            Grey solid lines indicate the contours of SFE of $10^{1}, 10^{0}$, and $10^{-1}\,\gyr^{-1}$.
        }
        \label{fig:ks_relation}
    \end{center}
\end{figure*}

We show the relationship between the area-averaged SFR, $\Sigma_\mathrm{SFR}$, and the gas surface density, $\Sigma_\mathrm{gas}$, \citep[so-called the Kennicutt-Schmidt (KS) relation,][]{schmidt_1959, kennicutt_1998} at $t=500\,\myr$ in Fig.~\ref{fig:ks_relation}.
The $\Sigma_\mathrm{SFR}$ and $\Sigma_\mathrm{gas}$ values are estimated in $750\,\pc \times 750\,\pc$ pixels, the sizes of which are comparable to the observational and theoretical studies \citep[e.g.][]{bigiel_2008, fujimoto_2019}.
In these plots, we calculate SFRs by counting star particles younger than $10\,\myr$ since this time-scale is comparable to that used in observations to determine SFR \citep[e.g.][]{koda_2006,kennicutt_2007}.
We fit the results on $\Sigma_{\mathrm{SFR}}$ as a function of $\Sigma_{\mathrm{gas}}$ with a power-law relation $\Sigma_{\mathrm{SFR}} \propto \Sigma^N_{\mathrm{gas}}$.

The top panels in Fig.~\ref{fig:ks_relation} show the KS relation for the SFRs calculated using all young stars.
We find that the power-law index, $N$, for the Coll simulation, $N=5.31\pm0.26$, is slightly larger than that for the Const simulation ($N=5.70\pm0.28$).
Comparing our simulated galaxies with the best-fit observational result from \citep{bigiel_2008}, the $N$ values are significantly higher in both simulations, probably due to the fact that our data points are mostly distributed around the threshold surface density of $\Sigma_\mathrm{gas} \simeq 9 \, \Msun$ \citep{bigiel_2008}. 

To investigate what makes the KS relation for Coll steeper than for Const, we show the KS relation with SFRs calculated using young stars born in colliding GMCs in the middle panels of Fig.~\ref{fig:ks_relation}.
In these plots, the power-law indices for both simulations are much higher than the cases of all young stars: $N=7.85\pm0.87$ for Const and $N=6.37\pm0.52$ for Coll.
Although the value of $N$ for Coll is smaller than that for Const, there are data with higher $\Sigma_\mathrm{SFR}$ in Coll than that in Const for a given $\Sigma_\mathrm{gas}$, especially at $\Sigma_\mathrm{SFR} \gtrsim 10^{-2.5}\,\Msun\,\yr^{-1}\,\kpc^{-2}$.
Therefore, the star formation efficiency (SFE) defined as $\Sigma_\mathrm{SFR}/\Sigma_\mathrm{gas}$  for the stars born in the colliding GMCs is higher in Coll than in Const as expected from the star formation model employed in Coll (Eq.~\ref{eq:epsilon_coll}). 

As shown in the bottom panels of Fig.~\ref{fig:ks_relation}, the KS relations by young stars born outside the colliding GMCs produce relatively lower $N$: $N=4.82\pm0.23$ for Const and $N=4.43\pm0.23$ for Coll.
The KS relations encompassing all young stars (shown in the top panels) are a synthesis of the KS relations attributed to stars born in the colliding GMCs and those born outside them (middle and bottom panels). 
As a result, the simulation with enhanced star formation within colliding GMCs shows elevated $N$ values, due to the substantial contribution from star formation activities within these colliding GMCs.

\subsection{GMCs}
\label{result:gmc}

\begin{figure*}
    \begin{center}
        \includegraphics[width=2\columnwidth]{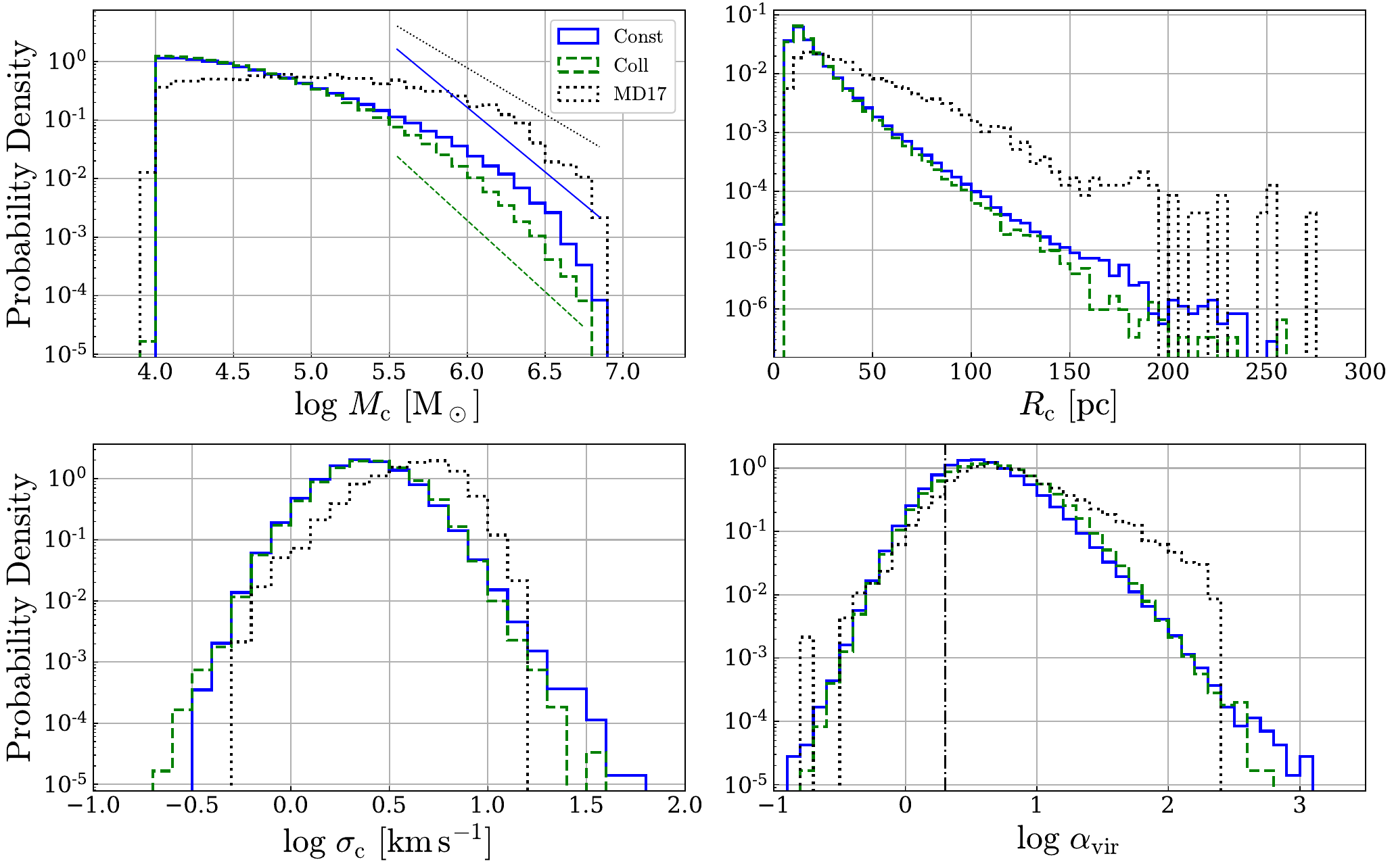}
        \caption{
            Probability density functions (PDFs) of physical properties of GMCs identified in snapshots from $t=350\,\myr$ to $500\,\myr$.
            The GMC properties for the Const and Coll simulations are shown by the blue solid and green dashed lines, respectively.
            We show the PDFs of the masses $M_\mathrm{c}$ (top left), the radii $R_\mathrm{c}$ (top right), the 1D velocity dispersion $\sigma_\mathrm{c}$ (bottom left), and the virial parameters $\alpha_\mathrm{vir}$ (bottom right) of GMCs.
            We also present the observational results of MW clouds \citep[][MD17]{miville-deschenes_2017} with the black dotted line in each panel for comparison.
            We plot only the MW clouds in the same mass range as our simulated GMCs for a consistent comparison.
            The thin lines in the $M_\mathrm{c}$ panel represent the power laws of $dN_\mathrm{c}/d\log M_\mathrm{c} \propto M_\mathrm{c}^{-\beta}$ with $\beta=2.20\pm0.17$ for Const and $\beta=2.43\pm0.11$ for Coll, and $\beta=1.59\pm0.16$ for the MW clouds in the regime of $M_\mathrm{c} > 10^{5.5}\,\Msun$.
            The black dash-dotted line in the $\alpha_\mathrm{vir}$ panel indicates $\alpha_\mathrm{vir}=2$ which is the boundary of whether GMCs are gravitationally bound or not. 
        }
        \label{fig:gmc:properties_1Dpdf}
    \end{center}
\end{figure*}

\subsubsection{GMC properties}
\label{result:gmc:gmc_properties}

We here show how the star formation model of CCCs affects the physical properties of GMCs.
Fig.~\ref{fig:gmc:properties_1Dpdf} shows the probability density functions (PDFs) of the GMC properties.
The GMCs are identified in $1\,\myr$ interval snapshots from $t=350-500\,\myr$ using the FoF algorithm as used in the on-the-fly identification of CCCs.
We also add observational results of MW clouds \citep[][]{miville-deschenes_2017} for comparison in Fig~\ref{fig:gmc:properties_1Dpdf}, where we exclude the MW clouds that have masses outside the range of our simulated GMC masses and that are in $2\,\kpc<R_\mathrm{gal}<13\,\kpc$ to make a consistent comparison between our simulation results and the observation.
We first show the PDF of GMC masses, $M_\mathrm{c}$, in the top left panel of Fig.~\ref{fig:gmc:properties_1Dpdf}.
For $M_\mathrm{c} \lesssim 10^5\,\Msun$, the distributions in our simulations are almost identical to each other.
We cannot track GMCs with $M_\mathrm{c} \lesssim 10^4\,\Msun$ due to the number threshold employed in the FoF grouping algorithm.
The Const and Coll simulations produce GMCs reaching at $M_\mathrm{c} \approx 8\times10^6\,\Msun$ and $\approx 6\times10^6\,\Msun$, respectively.
We fit the distribution functions with
\begin{equation}
    \frac{dN_\mathrm{c}}{d\log M_\mathrm{c}} \propto M_\mathrm{c}^{-\beta},
\end{equation}
where $N_\mathrm{c}$ is the number of GMCs contained in a given mass bin and $\beta$ is the index describing how GMC masses are distributed in the high-mass regime.
We use GMCs with $M_\mathrm{c} > 10^{5.5}\,\Msun$ for this fit, as in \citet{pettitt_2018}.
We find that the values of the power-law index, $\beta$, are $2.20\pm0.17$ for Const and $2.43\pm0.11$ for Coll.
The slightly higher value for the latter suggests that GMCs are difficult to grow when stars are effectively formed via CCCs.
In the Coll simulation, the escalated efficiency of star formation within colliding GMCs may result in more potent stellar feedback compared to the Const simulation. 
This intensified feedback could potentially initiate the disruption of the hosting GMC and impede the process of mass accumulation.
This result is consistent with \citet{li_2020}, who showed that using higher $\epsilon_\mathrm{ff,SF}$ in galaxy simulations leads to a steeper GMC mass function. 
While there are some differences between our simulations and \citet{li_2020} (e.g. simulation setup and GMC identification methods), our result is in line with theirs in terms of a steeper slope at the high-mass end by assuming a higher efficiency of star formation.

Compared with the observational results, less massive GMCs are more likely to exist in both simulations, while the observed clouds are the most probable at $M_\mathrm{c}\approx10^5\,\Msun$ as shown in the top left panel of Fig.~\ref{fig:gmc:properties_1Dpdf}. 
There are several reasons for this discrepancy.
First, our GMC identification process operates in three-dimensional space, as against the two-dimensional method used in the observations. 
As shown in \citet{grisdale_2018}, this difference in identification methods may lead to the identification of different populations of GMCs.
Furthermore, it is possible that physics not included in our simulations, such as magnetic fields, play an important role in shaping the GMC mass function. 

As is the GMC mass distribution, the fraction of the larger GMCs in their radii, $R_\mathrm{c}$, tends to be smaller in Coll than in Const (top right panel).  
In the range of $R_\mathrm{c} \lesssim 50\,\pc$, there are no significant differences between the Const and the Coll simulations with the median values of $\sim 15\,\pc$ for both.
These values are about half of the MW clouds observed by \citet{miville-deschenes_2017}.
Above this range, the fraction in Coll is smaller than that in Const, extending to $\sim 250\,\pc$ in both simulations.
Again the observation suggests flatter distribution than our simulations. 

There are only slight differences between our simulations in terms of the 1D velocity dispersion, $\sigma_\mathrm{c}$, and the virial parameter, $\alpha_\mathrm{vir}$, shown in the bottom panels of Fig.~\ref{fig:gmc:properties_1Dpdf}.
In the distributions of $\sigma_\mathrm{c}$, both simulations produce similar typical values of $\sim 2.5\,\kms$.
The fraction of GMCs with $\gtrsim 10\,\kms$ is slightly lower in Coll than in Const.
The values of $\alpha_\mathrm{vir}$ range from $\sim 0.1$ to $\sim 1000$, independent of the star formation models.
Only $17$ per cent and $14$ per cent of the GMCs are gravitationally bound (i.e. $\alpha_\mathrm{vir} < 2$) in Const and Coll, respectively.
Both simulations have a peak value of $\sim 4$, which is in agreement with \citet{miville-deschenes_2017}.
In summary, the internal kinematics of the GMCs are not strongly affected by the star formation models for the CCCs.

\subsubsection{Star formation efficiency per free-fall time}
\label{result:gmc:sfe_per_tff}

\begin{figure}
    \begin{center}
        \includegraphics[width=\columnwidth]{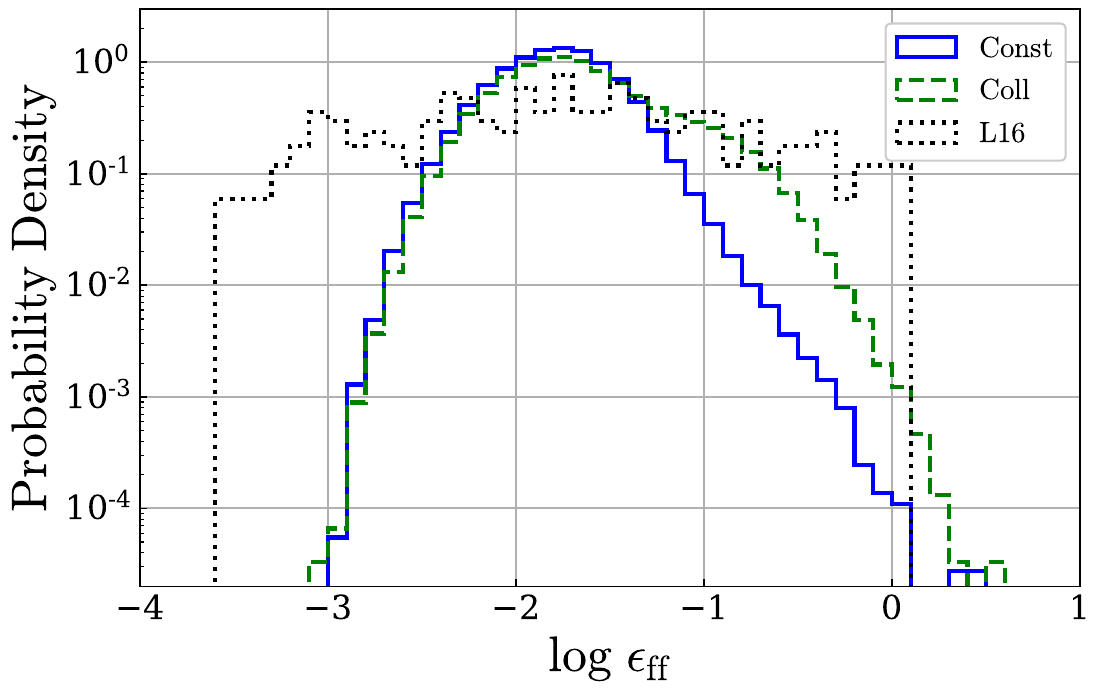}
        \caption{
            PDFs of the star formation efficiencies per free-fall time, $\epsilon_\mathrm{ff}$.
            The values of $\epsilon_\mathrm{ff}$ are calculated by grouping neighbouring dense gas elements and star particles with ages $<4\,\myr$ with the FoF.
            We also add the observational result of the MW clouds \citep[][L16]{lee_2016}.
        }
        \label{fig:gmc:sfe_1Dpdf}
    \end{center}
\end{figure}

We calculate the star formation efficiency per free-fall time, $\epsilon_\mathrm{ff}$, by running the FoF using both dense gas elements and star particles younger than $t_\mathrm{\star,y} = 4\,\myr$ in the simulation snapshots.
This choice of $t_\mathrm{\star,y}$ is comparable to the lifetime of young stars and is consistent with the previous studies of $\epsilon_\mathrm{ff}$ in both simulations and observations \citep[e.g.][]{lee_2016, grisdale_2021}.
The value of $\epsilon_\mathrm{ff}$ is computed with
\begin{equation}
    \epsilon_\mathrm{ff} = \frac{M_\mathrm{\star,y}}{M_\mathrm{\star,y}+M_\mathrm{c}} \frac{t_\mathrm{ff,c}}{t_\mathrm{\star,y}},
\end{equation}
where $M_\mathrm{\star,y}$ is the total mass of young stars and $t_\mathrm{ff,c} = \sqrt{3\pi/32G\bar{\rho}_\mathrm{c}}$ is the free-fall time of GMCs, assuming that GMCs are spheres of the uniform density of $\bar{\rho}_\mathrm{c} = 3M_\mathrm{c}/4\pi R_\mathrm{c}^3$.
We note again that $\epsilon_\mathrm{ff}$ is different from $\epsilon_\mathrm{ff,SF}$: the former is estimated for star-forming GMCs, and the latter is a parameter governing star formation in a fluid element \citep[see also][]{grisdale_2018}. The value of $\epsilon_\mathrm{ff}$ can be quite different from $\epsilon_\mathrm{ff,SF}$ due to feedback and other factors.
We show the PDF of the star formation efficiencies per free-fall time, $\epsilon_\mathrm{ff}$, in Fig.~\ref{fig:gmc:sfe_1Dpdf}, where the observational results of the MW clouds \citep[][]{lee_2016} are also included.
As well as in Fig.~\ref{fig:gmc:properties_1Dpdf}, we plot only the observational data in the mass range of our simulated GMCs in the disc region for analysis.
Both simulations produce a wide range of $\epsilon_\mathrm{ff}$ from $\sim 10^{-3}$ to an order of unity.
For $\epsilon_\mathrm{ff} \lesssim 3\times10^{-2}$, the distributions are similar to each other, while the Coll simulation produces only a slightly smaller fraction.
On the other hand, for $\epsilon_\mathrm{ff} \gtrsim 3\times10^{-2}$, the probability densities in Coll are $\sim0.1-1.3$ dex larger than those in Const.
This fact shows that a larger fraction of GMCs efficiently forms stars as a result of the enhanced star formation in CCCs in the Coll simulation.

Despite the difference in the higher value of $\epsilon_\mathrm{ff}$, the median values are comparable to each other.
The median for Const is $\sim 1.7\times10^{-2}$ and that for Coll is $\sim 2.0\times10^{-2}$, which are consistent with MW clouds \citep[$1.8\times10^{-2}$,][]{lee_2016} and extragalactic GMCs \citep[$3\times10^{-3}-2.6\times10^{-2}$,][]{utomo_2018}.
We calculate the standard deviation from the median absolute deviation of $\log\epsilon_\mathrm{ff}$, $\sigma_{\log\epsilon_\mathrm{ff}}$, as in \citet[][see also \citet{muller_2000}]{grisdale_2021}.
We find that the values of $\sigma_{\log\epsilon_\mathrm{ff}}$ are $\sim 0.30$ and $\sim 0.38$ in the Const and Coll simulations, respectively and thus the distribution in the latter is broader than the other.
However, the $\epsilon_\mathrm{ff}$ distributions in our simulations are narrower than MW clouds.
The value of $\sigma_{\log\epsilon_\mathrm{ff}}$ for the MW clouds\footnote{We measure this value from \citet{lee_2016}, using the data with masses comparable to our GMCs.} is $\sim 0.85$, which is more than twice as large as our simulation results.
This discrepancy is due to the fact that our simulations do not have very inefficient GMCs ($\epsilon_\mathrm{ff} < 10^{-3}$), while \citet{lee_2016} observed such inefficient clouds.
To explain the GMCs with such low SFE, we probably need to further improve our star formation model. 

\subsubsection{Cloud collision speeds}
\label{result:gmc:v_coll}

\begin{figure}
    \begin{center}
        \includegraphics[width=\columnwidth]{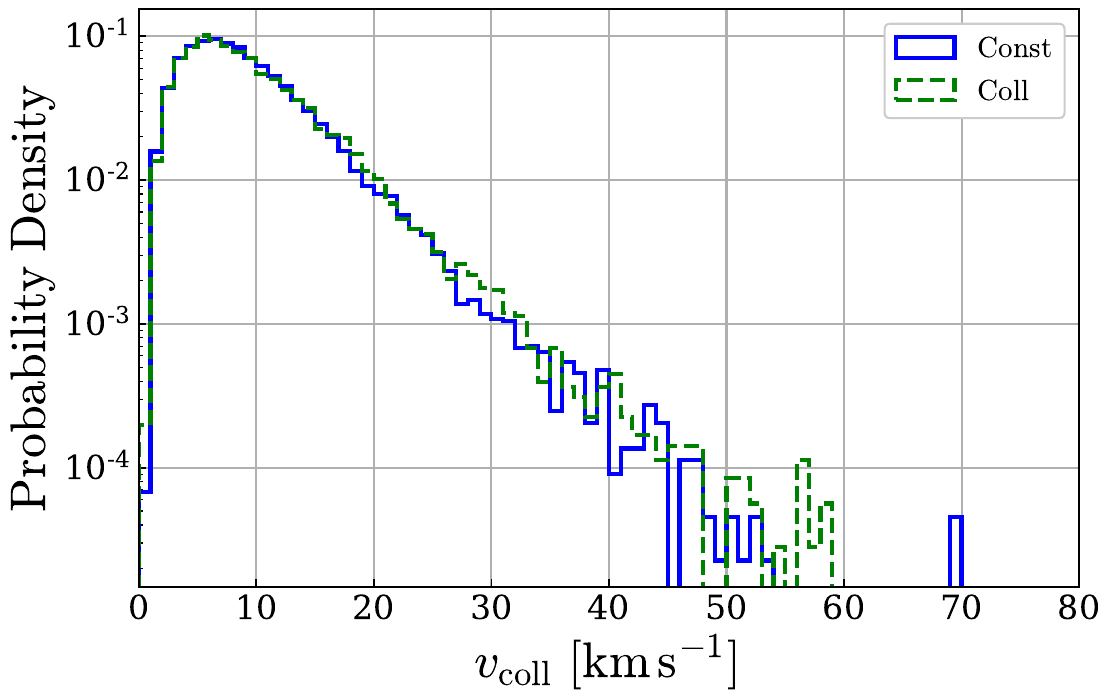}
        \caption{
            PDFs of the collision speed between GMCs, $\varv_\mathrm{coll}$.
            Each collision event is identified with the on-the-fly algorithm.
        }
        \label{fig:gmc:vcoll_1Dpdf}
    \end{center}
\end{figure}

Since our star formation model for CCCs depends on the cloud collision speed, $\varv_\mathrm{coll}$, and some numerical simulations of CCCs suggest its importance for core formation, it is essential to investigate the value of $\varv_\mathrm{coll}$.
In Fig.~\ref{fig:gmc:vcoll_1Dpdf}, we show the PDF of the cloud collision speed, $\varv_\mathrm{coll}$, identified by the on-the-fly CCC finder.
The distributions of $\varv_\mathrm{coll}$ do not differ significantly between our simulations, suggesting that star formation triggered by CCCs does not affect collision speeds.
About $65$ per cent of collisions occur at relative speeds less than $10\,\kms$, with peaks at $\sim 7\,\kms$.
Beyound the peaks, the probabilities decreases with increasing $\varv_\mathrm{coll}$, reaching $\sim 70\,\kms$ in Const and $\sim 59\,\kms$ in Coll.

\begin{figure*}
    \begin{center}
        \includegraphics[width=2\columnwidth]{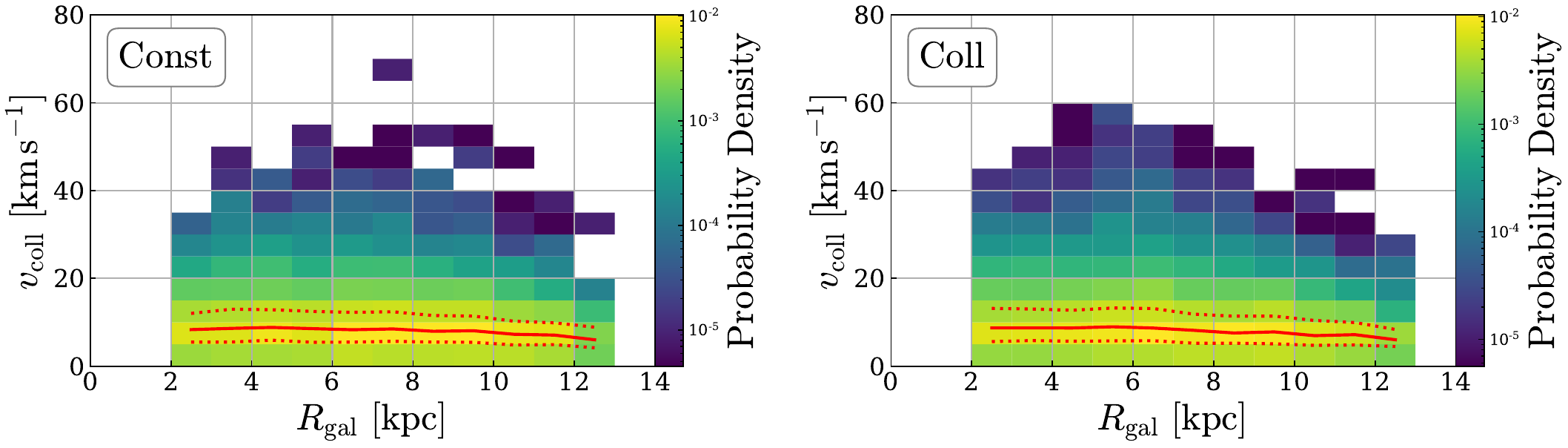}
        \caption{
            2D PDFs of the collision speed, $\varv_\mathrm{coll}$, and the galactic radii, $R_\mathrm{gal}$, for the Const (left) and Coll (right) simulations.
            The red solid and dotted lines denote the median values, 25 and 75 percentiles of $\varv_\mathrm{coll}$ in a given annulus with $1\,\kpc$ width.
         }
        \label{fig:gmc:vcoll_galrad_2Dpdf}
    \end{center}
\end{figure*}

We also show the collision speed as a function of galactic radius in Fig.~\ref{fig:gmc:vcoll_galrad_2Dpdf}.
There are apparently no significant differences between the simulations in this plot.
The most probable $\varv_\mathrm{coll}$ in a given annulus with $1\,\kpc$ width is between $5$ and $10\,\kms$ regardless of the galactic radius.
We find that the median values of $\varv_\mathrm{coll}$ become slightly smaller as the radius increases, suggesting that cloud collision speeds are weakly dependent on the galactic rotation.

\begin{figure*}
    \begin{center}
        \includegraphics[width=2\columnwidth]{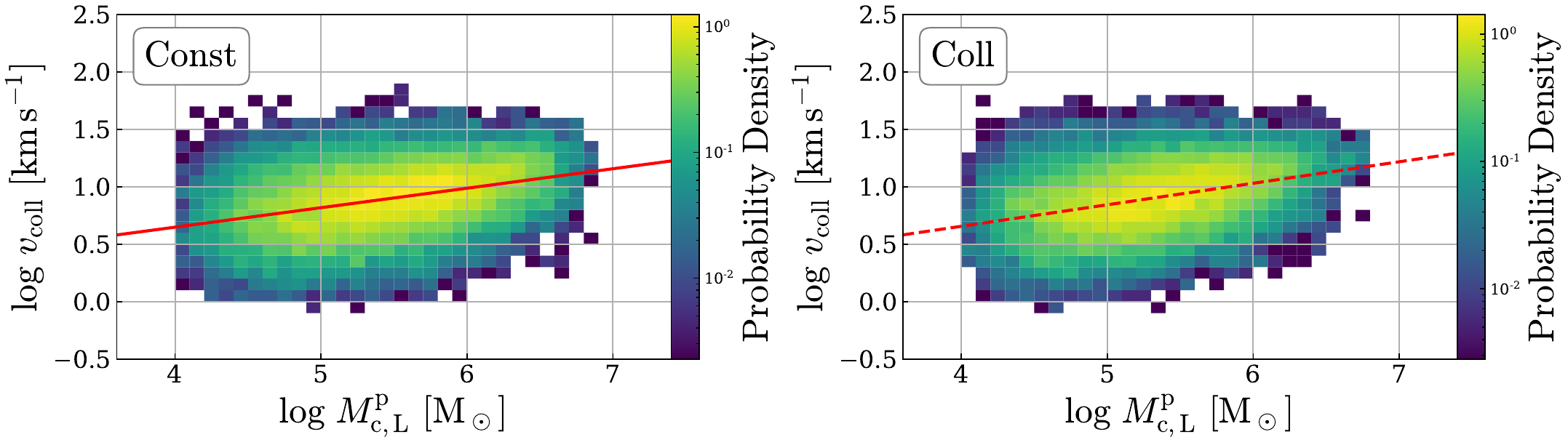}
        \caption{
            2D PDFs of the collision speed, $\varv_\mathrm{coll}$, as a function of the mass of the most massive GMC progenitor just before the collision, $M_\mathrm{c,L}^\mathrm{p}$, for the Const (left) and Coll (right) simulations.
            The red line shows the best fit of $\varv_\mathrm{coll} \propto \left( M_\mathrm{c,L}^\mathrm{p} \right)^a$ with $a = 0.19\pm0.0022$ for Const and $a = 0.21\pm0.0027$ for Coll.
        }
        \label{fig:gmc:vcoll_Mc_2Dpdf}
    \end{center}
\end{figure*}

In Fig.~\ref{fig:gmc:vcoll_Mc_2Dpdf}, 
we show the collision speeds as a function of the mass of the most massive GMC progenitor at one previous timestep before the collision, $M_\mathrm{c,L}^\mathrm{p}$. 
We find that $\varv_\mathrm{coll}$ is weakly correlated with $M_\mathrm{c,L}^\mathrm{p}$. We fit the distribution by the function
\begin{equation}
    \varv_\mathrm{coll} \propto \left( M_\mathrm{c,L}^\mathrm{p} \right)^a,
\end{equation}
where $a = 0.19\pm0.0022$ for the Const and $a = 0.21\pm0.0027$ for the Coll.
Although the correlation is weak, with the correlation coefficient of $r=0.37$ for both, this result implies that more massive GMCs tend to collide at higher collision speeds. 
The power-law index in the correlation is less than $a=0.5$ which is obtained by assuming that clouds collide at free-fall speeds between two clouds \citep[see][]{fujimoto_2020}.
In our simulations, the dependence on the GMC masses is weakened, possibly due to the presence of stellar feedback. 

The ranges of collision speeds are consistent with observational reports \citep[see Table~1 in][and references therein]{fukui_2021}.
\citet{finn_2019} observed a collision with $\varv_\mathrm{coll} > 100\,\kms$ in the merging Antennae galaxies, and \citet{fujimoto_2014_b} simulated a barred galaxy and found $\varv_\mathrm{coll} > 100\,\kms$ in the bar region.
However, we cannot reproduce such a high-speed collision in our simulations of isolated disc galaxies without imposed spiral arms and bars.
These discrepancies imply that collision speeds depend on galactic environments such as galactic structures and galaxy mergers.

\subsubsection{Cloud collision frequency}
\label{result:gmc:f_coll}

\begin{figure}
    \begin{center}
        \includegraphics[width=\columnwidth]{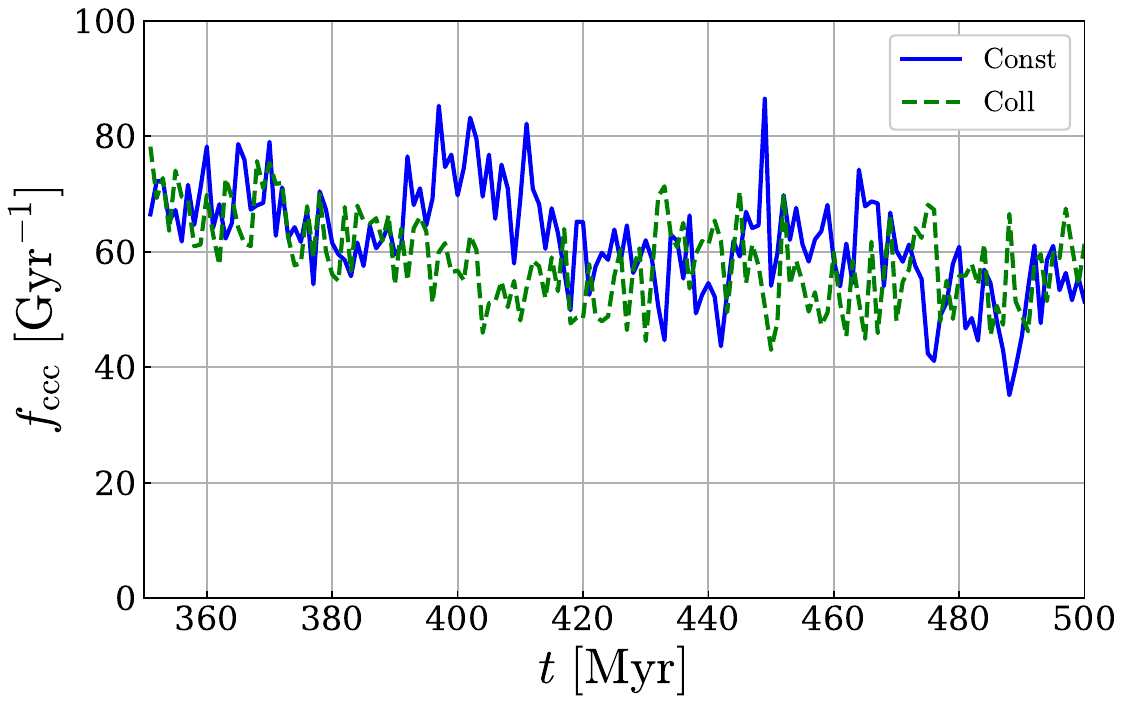}
        \caption{
            Cloud collision frequency, $f_\mathrm{ccc}$, as a function of time.
        }
        \label{fig:gmc:collision_frequency}
    \end{center}
\end{figure}

\begin{table}
    \caption{
        The average number of GMCs estimated from snapshots, 
        the total number of CCC events over the tracking time $\Delta T_\mathrm{track} = 150\,\myr$, 
        the average cloud collision frequency in units of $\gyr^{-1}$
        for each simulation.
    }
    \label{tab:average_fcoll}
    \begin{center}
	\begin{tabular*}{0.99\columnwidth}{@{\extracolsep{\fill}}llcc}
        \hline
            Values & Meaning & Const & Coll\\
            \hline
            $\bar{n}_\mathrm{c}$ & average number of GMCs & $4721$ & $4015$\\[2pt]
            $n_\mathrm{ccc}$ & total number of collisions & $44096$ & $35470$\\[2pt]
            $\bar{f}_\mathrm{ccc}\,[\gyr^{-1}]$ & average collision frequency & $62.3$ & $58.9$\\[2pt]
            \hline
        \end{tabular*}
    \end{center}
\end{table}

The frequency of CCCs is also one of the key factors for star formation in colliding GMCs.
We here present the cloud collision frequencies as a function of time $t$ in Fig.~\ref{fig:gmc:collision_frequency}.
The frequency, denoted as $f_\mathrm{ccc}$, characterizes the rate at which a GMC engages in collisions with others over a specified interval of time:
\begin{equation}
    f_\mathrm{ccc}(t) = \frac{n_\mathrm{ccc}^{\Delta T}(t)}{n_\mathrm{c}(t-\Delta T)\,\Delta T},
    \label{eq:fcoll_time}
\end{equation}
where $\Delta T = 1\,\myr$ is the time interval between snapshots, $n_\mathrm{ccc}^{\Delta T}(t)$ is the number of collisions recorded by the on-the-fly algorithm between $t-\Delta T$ and $t$ (i.e. how many times collisions occur between snapshot outputs), and $n_\mathrm{c}(t-\Delta T)$ is the number of GMCs at time $t-\Delta T$ (i.e. the number of GMCs in one previous snapshot).
Significant fluctuations in cloud collision frequencies are evident in both simulations, with variations spanning approximately $10-25\,\gyr^{-1}$ over short temporal intervals lasting a few million years ($\myr$). These variations amount to roughly $15-40$ per cent of the mean collision frequencies, which are approximately $60\,\gyr^{-1}$.
These fluctuations in collision frequencies are closely tied to the corresponding variations in SFRs, as illustrated in Fig.~\ref{fig:sfh_cmp_coll}.
The variation in collision frequency is found to be larger in the Const case than in the Coll case. 
This may be due to the higher star formation efficiency in the Coll case and the resulting stronger stellar feedback. 
These factors combine to reduce the number of GMCs and the likelihood of collisions between GMCs.

We also compute the mean collision frequency, $\bar{f}_\mathrm{ccc}$, in a similar way to Eq.~(\ref{eq:fcoll_time}) as
\begin{equation}
    \bar{f}_\mathrm{ccc} = \frac{n_\mathrm{ccc}}{\bar{n}_\mathrm{c}\,\Delta T_\mathrm{track}},
    \label{eq:average_fcoll}
\end{equation}
where $\Delta T_\mathrm{track} = 150\,\myr$ is the tracking time in our analysis, $n_\mathrm{ccc}$ is the total number of collisions over $\Delta T_\mathrm{track}$, $\bar{n}_\mathrm{c}$ is the average number of GMCs estimated from the number of GMCs in snapshots \citep[see also][]{fujimoto_2020}.
These values are listed in Table~\ref{tab:average_fcoll}.
Both simulations have comparable average collision frequencies. While the Coll simulation has a lower average number of GMCs than the Const simulation, the total number of collisions in Coll is also lower than in Const, resulting in similar collision frequencies.
Stellar feedback is often considered to be the primary driver of cloud collisions \citep[e.g.][]{skarbinski_2023}. 
Slightly enhanced feedback in Coll probably maintains the collision frequency even with the smaller number of GMCs.
These values are $1.5-2$ higher than the estimates from the post-processing of galaxy simulations
\citep[$30-40\,\gyr^{-1}$,][]{tasker_2009, tasker_2011, dobbs_2015}.
We discuss this in detail in Sec.~\ref{discussion:ccc_identification}.

\subsubsection{Masses of colliding clouds}
\label{result:gmc:ccc_event}

\begin{figure*}
    \begin{center}
        \includegraphics[width=2\columnwidth]{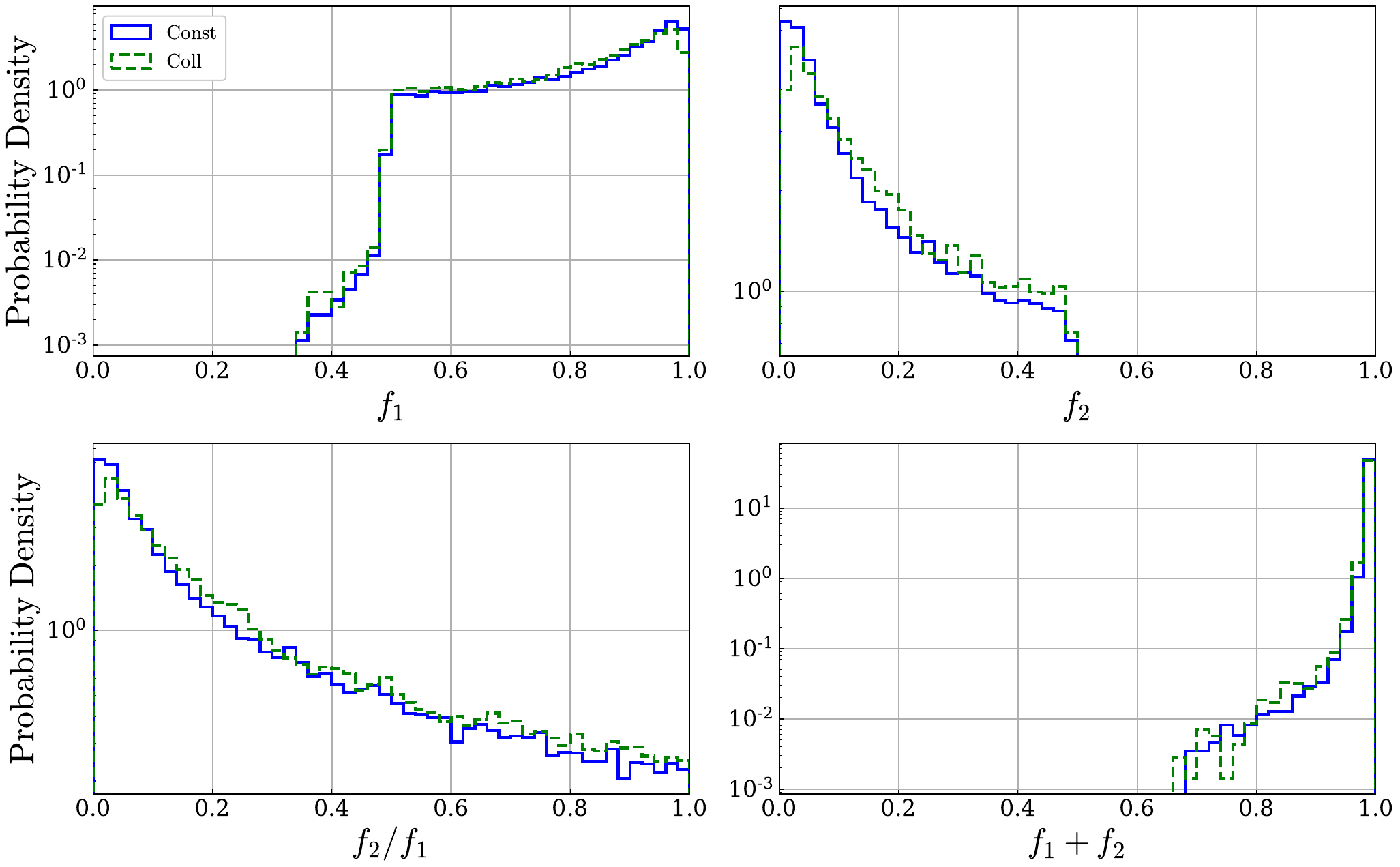}
        \caption{
            Top left: PDF of mass fractions contributing most to colliding GMCs, $f_1$.
            Top right: PDF of mass fractions which is the second largest contributor to colliding GMCs, $f_2$.
            Bottom left: PDF of ratios of $f_2$ to $f_1$.
            Bottom right: PDF of sums of $f_1$ and $f_2$.
        }
        \label{fig:gmc:collision_fraction_1Dpdf}
    \end{center}
\end{figure*}

We here explore the masses and mass ratios of colliding clouds in our simulations.
We compute the fraction of mass derived from an individual progenitor within a colliding GMC. The most significant and second most significant fractions are denoted as $f_1$ and $f_2$, respectively as we described in Sec.~\ref{method:CCC_identification}. 
The top left and top right panels of Fig.~\ref{fig:gmc:collision_fraction_1Dpdf} show the PDFs of the mass fractions $f_1$ and $f_2$ respectively.
The median of $f_1$ is $0.89$ for the Const simulation and $0.86$ for the Coll simulation. The distribution of $f_2$ is skewed towards very small values, with medians of $0.11$ and $0.13$ for the Const and Coll simulations respectively.
These results suggest that most of the mass of a colliding cloud comes from a single progenitor.
We also find that these distributions show minimal dependence on the chosen star formation model. 

\begin{figure*}
    \begin{center}
        \includegraphics[width=2\columnwidth]{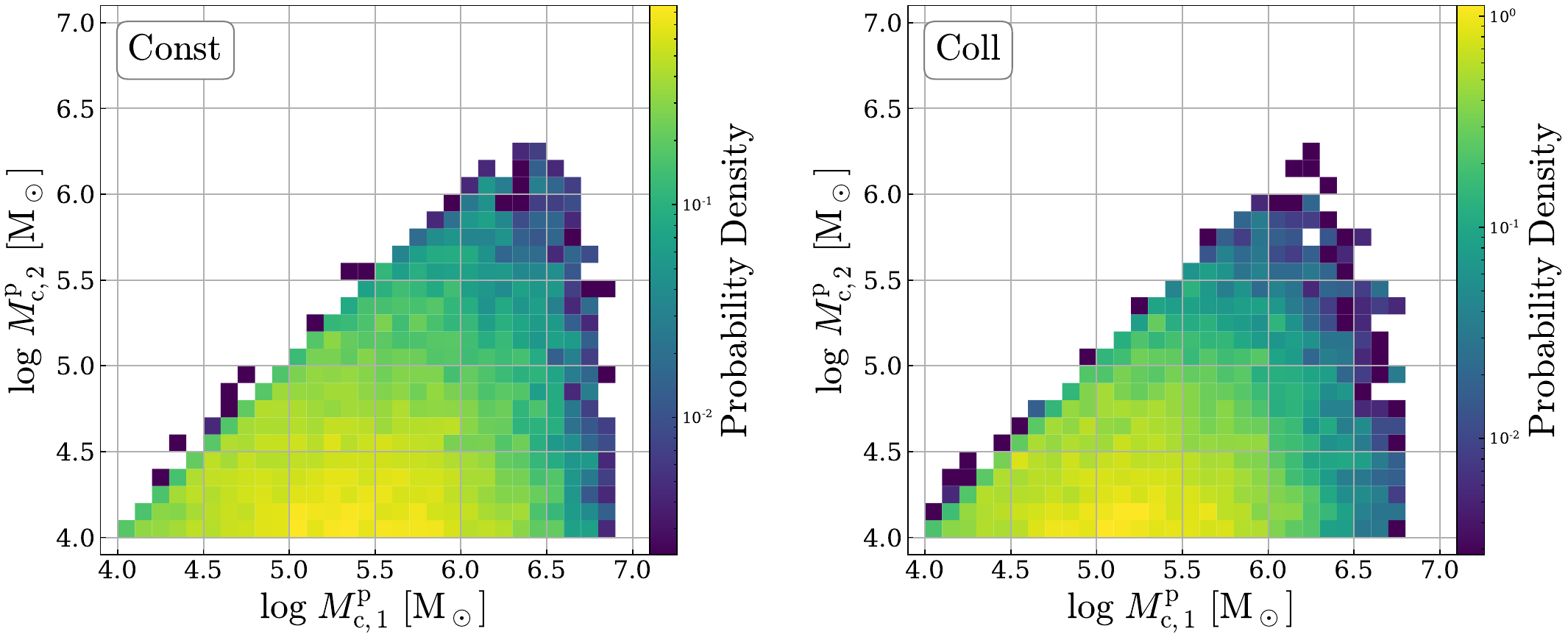}
        \caption{
            2D PDFs of CCCs as a function of masses of GMC pairs just before collisions for the Const (left) and Coll (right) simulations.
            The GMC masses, denoted as $M_\mathrm{c,1}^\mathrm{p}$ and $M_\mathrm{c,2}^\mathrm{p}$, represent the masses of progenitor GMCs primarily contributing to the colliding GMC mass and the next most significant contributor, respectively.
        }
        \label{fig:gmc:cloud_mass_before_collision_2Dpdf}
    \end{center}
\end{figure*}

For a more direct exploration of colliding mass ratios, we present the distribution of $f_2/f_1$ in the lower left panel of Fig.~\ref{fig:gmc:collision_fraction_1Dpdf}.
We find that the values of $f_2/f_1$ tend to be more probable at very low values and their median values are $\sim0.12$ for Const and $\sim0.15$ for Coll, suggesting that a smaller GMC typically collides with a $\gtrsim 7$ times massive GMC.
This is consistent with \citet{dobbs_2015}, who analysed cloud mergers in a simulated galaxy with a post-processing method.
Furthermore, our result aligns with the underlying assumption in simulations of CCCs that CCCs occur between clouds with a small mass ratio \citep[e.g.][]{habe_1992}.
However, the mass ratios in such simulations are usually much greater than $0.1$, which is slightly inconsistent with our typical collisions.
We lastly show the PDF of $f_1+f_2$ in the bottom right panel of Fig.~\ref{fig:gmc:collision_fraction_1Dpdf}.
Both distributions have a peak at $f_1 + f_2 \sim 1$, suggesting that CCCs generally occur between two GMCs.
The median values of $f_1+f_2$ are above $0.99$ in both simulations and the minimum values are $0.68$ for both.
This result would justify our modelling of the CCC as a two-body collision.

Next, we explore the distribution of progenitor masses of CCCs. The masses of the progenitor GMCs that have the largest and second largest contributions to a colliding GMC are represented by $M_{\mathrm{c}, 1}^\mathrm{p}$ and $M_{\mathrm{c}, 2}^\mathrm{p}$, respectively, as explained in Sec.~\ref{method:CCC_identification}. Note that $M_{\mathrm{c}, 1}^\mathrm{p} \ge M_{\mathrm{c}, 2}^\mathrm{p}$ does not necessarily hold in all cases.
In Fig.~\ref{fig:gmc:cloud_mass_before_collision_2Dpdf}, we show the 2D PDF of CCCs as a function of  $M_{\mathrm{c}, 1}^\mathrm{p}$ and $M_{\mathrm{c}, 2}^\mathrm{p}$.
We find that the most probable pre-collision GMC masses are $\sim 10^4\,\Msun$ and $\sim 10^5$ to $10^{5.5}\,\Msun$ in both simulations, corresponding to a smaller GMC colliding with $\sim 10-30$ times larger ones.
There are also collisions with a mass ratio greater than $100$.
We note that the most likely pre-collision secondary GMC mass, $M_{\mathrm{c, 2}}^\mathrm{p} \simeq 10^4\,\Msun$ is probably determined by the lower mass limit of the GMC that we have used. 
We may need to consider the effect of a collision between a much smaller GMC, for example $10^3 \, \Msun$, and a typical primary GMC with a mass of $\sim 10^5\,\Msun$. 
Although the studies on CCCs in cloud-scale simulations did not assume such very high mass ratios, such collisions often occur and are the main contributor to the triggered star formation in our simulated galaxies (see Fig.~\ref{fig:Mstar_vs_Mcloud_collision}).

\section{Discussion}
\label{discussion}

\subsection{Star formation model of cloud-cloud collisions}
\label{discussion:sf_model_of_ccc}

We have developed a star formation model tailored for CCCs within galaxy-scale simulations, based on the simulations of CCCs performed by \citet{takahira_2018}, as a first attempt to include the CCC-induced star formation in galaxy simulations.
This model takes the form of a functional relationship with collision speed that captures the numerical result that increasing collision speed corresponds to decreasing star formation efficiency.
Star formation in colliding GMCs in the Coll simulation contributes $\sim 70$ per cent to the total SFR, which is a few tens of per cent higher than the semi-analytical estimate of \citet{kobayashi_2018}.
This discrepancy could be originated from the methodological differences between our galaxy simulations and the semi-analytical model as well as the different assumptions applied to the CCC-driven star formation. 
The typical value of the star formation efficiency, $\epsilon_\mathrm{ff}$, in Coll is comparable to observations \citep[][]{lee_2016}, although the distribution of $\epsilon_\mathrm{ff}$ is narrower than that of MW clouds.
While our star formation model can produce results that are in general agreement with observational data, further refinement is needed to achieve better agreement with observations. 

Our model depends only on $\varv_\mathrm{coll}$ and does not consider other cloud properties, such as mass ratios, turbulence, density structures, morphologies of GMCs, etc. for simplicity.
\citet{liow_2020} simulated CCCs and found that the SFR in colliding clouds increases when the collision speed is faster, the clouds are denser, and the clouds are less turbulent.
Notably, the dependence on collision speed contrasts with the finding of \citet{takahira_2018}.
\citet{sakre_2021} investigated the formation of massive cores in colliding magnetized clouds and showed that the stronger magnetic field leads to a greater number of massive cores.
In addition, \citet{sakre_2023} showed that a magnetic field plays an important role in massive core formation depending on the duration of CCCs.
Numerical simulations of CCCs such as these studies will be helpful for making a better model of star formation triggered by CCCs in galaxy simulations.

\begin{figure}
    \begin{center}
        \includegraphics[width=\columnwidth]{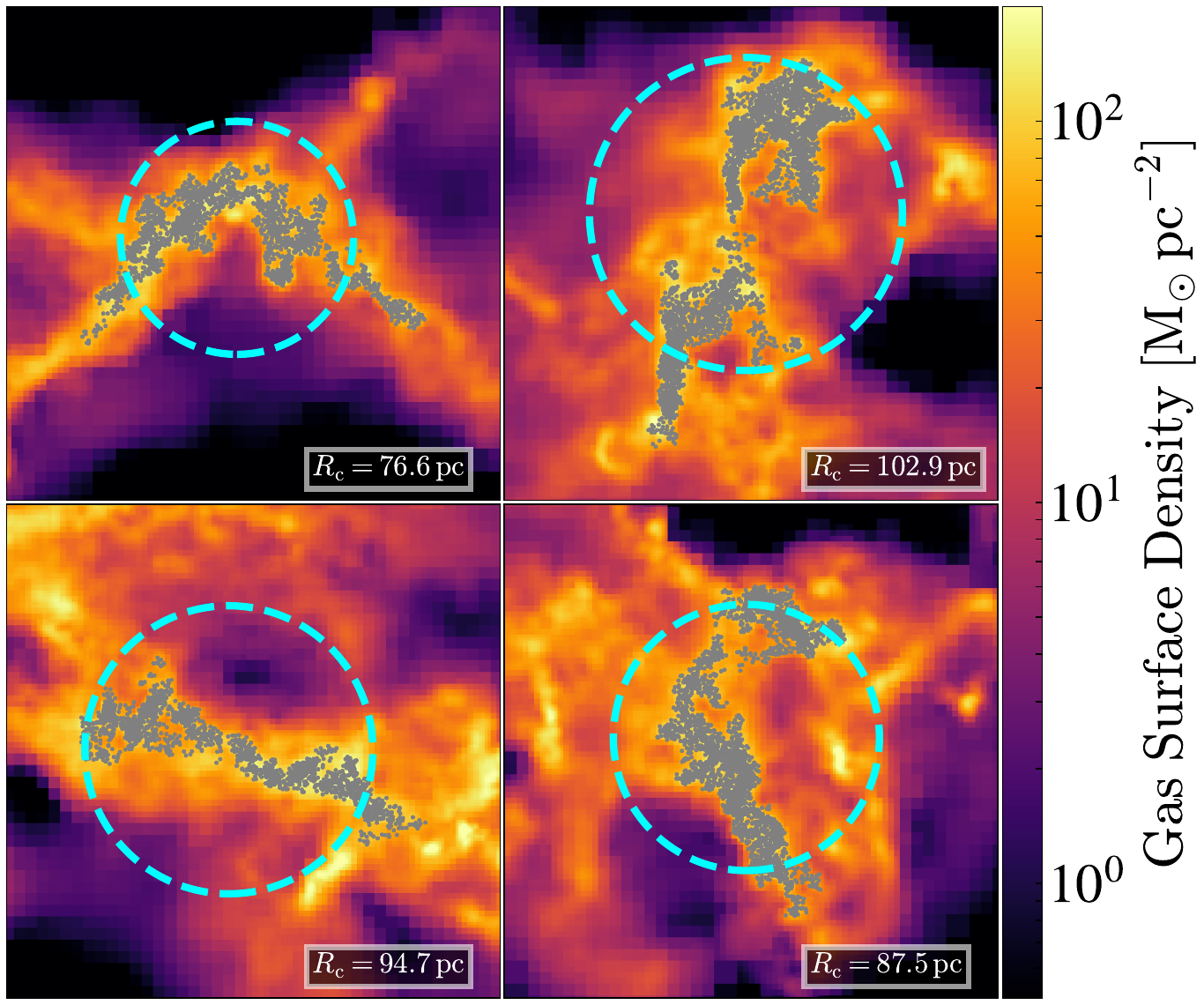}
        \caption{
            Examples of elongated GMCs in the Coll simulation.
            We show the zoomed-in gas maps around the GMCs in the face-on view of the galaxy with grey markers indicating the positions of gas elements that make up the GMCs superimposed. 
            The cyan dashed circles represent the GMCs when they are assumed to be spheres with their GMC radii, $R_\mathrm{c}$, centred on their centres of mass.
            Each value of $R_\mathrm{c}$ is shown at the bottom left of each panel.
        }
        \label{fig:elongated_clouds}
    \end{center}
\end{figure}

While GMCs in our simulations are sometimes elongated and filamentary as shown in Fig.~\ref{fig:elongated_clouds}, for simplicity, we assume that they are spherical. 
These elongated GMCs appear to exist in shell-like structures created by stellar feedback, since the GMCs in Fig.~\ref{fig:elongated_clouds} are close to regions of lower gas surface density.
In places close to SNe, gas condenses into GMCs and CCCs are expected to occur \citep[e.g.][]{inutsuka_2015,skarbinski_2023}.
\citet{li_2020} also showed that GMCs in galaxy simulations typically have ellipsoidal shapes, especially when stellar feedback is included.
However, star formation in colliding GMCs is likely to differ depending on whether a GMC collides along the minor axis or the major axis of an elongated one since the duration of a collision depends on the direction \citep[][]{abe_2022}.
Such a morphological effect on CCCs may need to be taken into account in our model.
The collision speed is a key parameter influencing the duration of these interactions. 
In the CCC simulations of \citet{takahira_2018}, which serve as the basis for our current CCC-induced star formation model, the collision speed is limited to a range of 5 to $30\,\kms$, but in our simulations, it extends beyond this range to include collision speeds of approximately $1$ to $70\,\kms$. 
To further our understanding and refinement of the CCC-triggered star formation model in the context of galaxy simulations, it would be beneficial to perform simulations with very high collision velocities, exceeding the $30\,\kms$ threshold.
Moreover, previous numerical studies of unequal-size cloud-cloud simulations often assume that a smaller cloud collides with a $<10$ times larger one.
Although we have modelled the star formation in colliding GMCs from the cloud-cloud simulations at these mass ratios, such ratios are lower than what we have found for the masses of GMC pairs just before collisions (Fig.~\ref{fig:gmc:cloud_mass_before_collision_2Dpdf}).
Star formation in colliding clouds with mass ratios greater than $10$ should also be investigated in order to improve the star formation model of CCCs.

\subsection{Post-processing vs on-the-fly identification of cloud-cloud collisions}
\label{discussion:ccc_identification}

\begin{figure*}
    \begin{center}
        \includegraphics[width=2\columnwidth]{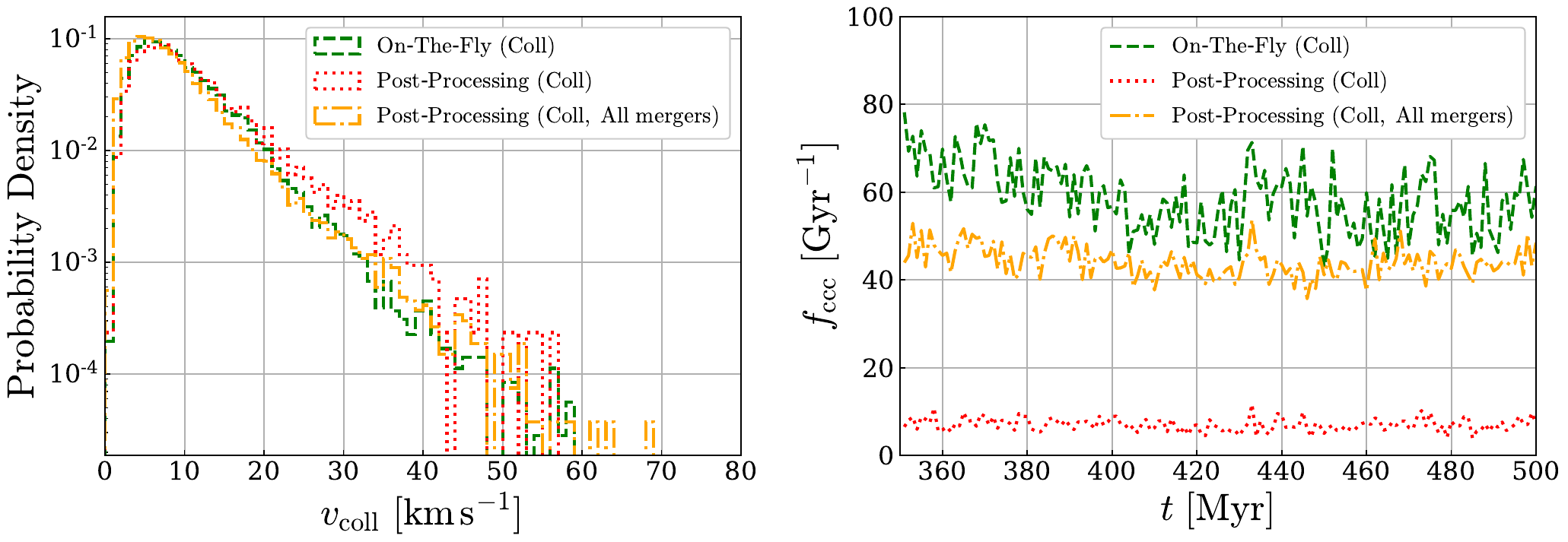}
        \caption{
            Properties of CCCs identified with the on-the-fly (green solid) and the post-processing (red dotted) analysis for the Coll simulation.
            We also plot the properties in the post-processing analysis using only the criterion~\ref{ccc_cond1} (orange dash-dotted) to present the contribution from all merging GMCs.
            The results from the on-the-fly analysis are the same as those shown in the Results section.
            Left: PDF of collision speeds, $\varv_\mathrm{coll}$.
            The value of $\varv_\mathrm{coll}$ for the post-processing analysis is estimated using the velocities of GMCs in one previous snapshot before collisions are found.
            Right: Cloud collision frequencies, $f_\mathrm{ccc}$, as a function of time.
            The value of $f_\mathrm{ccc}$ for the post-processing analysis is computed with Eq.~(\ref{eq:fcoll_time}) using the number of colliding GMCs, $n_\mathrm{coll}^{\Delta T}(t)$, instead of the number of collisions found by the on-the-fly method over the snapshot intervals, $n_\mathrm{ccc}^{\Delta T}$.
        }
        \label{fig:collision_vel_freq_otfppa}
    \end{center}
\end{figure*}

Previous studies of CCCs in galaxy simulations have analysed collision events using the information on gas elements in simulation snapshots \citep[e.g.][]{dobbs_2015}.
The separations of the simulation outputs in these simulations are typically $\sim 1\,\myr$.
In contrast to such post-processing analysis, we have developed the on-the-fly algorithm to identify CCCs, where we are able to find collisions with the time interval of the timestep $\Delta t \sim 100- 1000\,\yr$.
In this section, we investigate how the collision properties differ between the post-processing identification as used in the previous studies and the on-the-fly identification we have newly developed.
In other words, we analyse the dependence of the time intervals to find collisions on their properties.

\begin{figure*}
    \begin{center}
        \includegraphics[width=2\columnwidth]{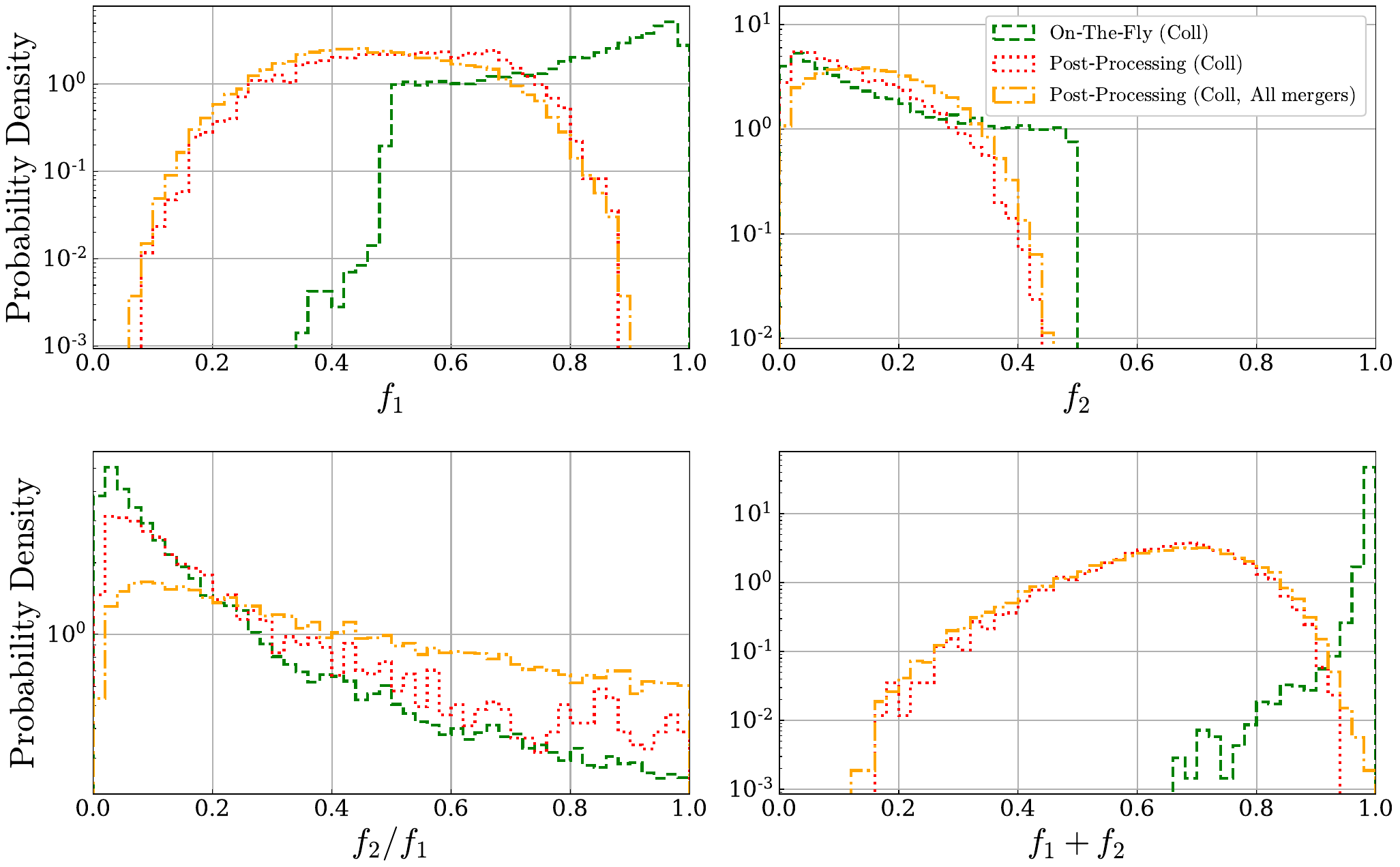}
        \caption{
            Same as Fig.~\ref{fig:gmc:collision_fraction_1Dpdf}, but for the Coll results analysed by the on-the-fly and the post-processing algorithms.
        }
        \label{fig:collision_frac_otfppa}
    \end{center}
\end{figure*}

The post-processing analysis of CCCs in this research is similar to the on-the-fly method described in Sec.~\ref{method:CCC_identification}:
we use GMCs and their compositing gas elements in simulation snapshots instead of those at each timestep and the same criteria for CCC identification.
Consequently, we study the differences between the timescales of $1\,\myr$ for the post-processing analysis and of $\sim 100- 1000\,\yr$ for the on-the-fly analysis.
We also present the results of CCC identification using solely the criterion \ref{ccc_cond1}, which is described in Sec.~\ref{method:CCC_identification}. Because prior numerical studies of CCCs in galaxy simulations have concentrated primarily on cloud mergers and ignored whether collisions can initiate star formation, we include these results in our post-processing analysis for comparison.
Unless otherwise stated, CCCs are identified by applying both criteria~\ref{ccc_cond1} and \ref{ccc_cond2} in the post-processing analysis.
We exclusively present the results for Coll, as no substantial distinctions were observed between Coll and Const regarding the distinction between the on-the-fly and post-process analyses.
The results for the on-the-fly analysis shown in Fig.~\ref{fig:collision_vel_freq_otfppa} and \ref{fig:collision_frac_otfppa} are the same as those shown in Sec.~\ref{result}.

In the left panel of Fig.~\ref{fig:collision_vel_freq_otfppa}, we show the PDFs of collision speeds, $\varv_\mathrm{coll}$.
The value of $\varv_\mathrm{coll}$ for the post-processing analysis is calculated from the relative velocity of GMCs in one previous snapshot before they are judged to have collided at a given snapshot.
The median values of $\varv_\mathrm{coll}$ do not differ significantly between the identification methods: $\sim 8.1\,\kms$ for the on-the-fly and $\sim 8.8\,\kms$ for the post-processing.
The maximum collision speed for the post-processing analysis is $\sim 57\,\kms$, which is comparable to that for the on-the-fly.

Despite these similarities, the probabilities for the post-processing are higher in the range of $\gtrsim 20\,\kms$ than for the on-the-fly analysis.
This difference is obviously caused by the different time intervals. 
Since the post-processing analysis cannot track a dynamical effect on the GMCs and changes in the bulk velocities of the GMCs during snapshot intervals, their collision speeds can be higher than those for the on-the-fly.
This result indicates that collision speeds determined by the post-processing analysis are slightly overestimated compared to the actual collision speeds.
Furthermore, we find that the collision speed tends to be slightly slower when all merging GMCs are taken into account in the post-processing analysis.
The median collision speed is $\sim7.1\,\kms$ and the distribution is slightly shifted to lower values compared to the others, especially in $\varv_\mathrm{coll}\lesssim20\,\kms$.

The right panel of Fig.~\ref{fig:collision_vel_freq_otfppa} shows the cloud collision frequencies, $f_\mathrm{ccc}$, as a function of time.
In order to calculate $f_\mathrm{ccc}$ in post-processing analysis, we identify the colliding GMCs using snapshots at time $t$ and its preceding snapshot.
There is a significant difference in collision frequencies between the on-the-fly and post-processing analyses. 
While the frequencies for the on-the-fly analysis are $\sim 40-80\,\gyr^{-1}$, those for the post-processing analysis are $\lesssim 10\,\gyr^{-1}$.
We compute the average collision frequency, $\bar{f}_\mathrm{ccc}$, for the post-processing analysis with Eq.~(\ref{eq:average_fcoll}) using the total number of colliding GMCs instead of the total number of collisions, $n_\mathrm{ccc}$.
This gives us $\bar{f}_\mathrm{ccc}\approx7.1\,\gyr^{-1}$, which is $\sim1/8$ of the on-the-fly analysis, showing that the post-processing misses a significant number of collision events.

While we find that the average collision frequency in the on-the-fly analysis is approximately $1.5-2$ times higher than the frequencies reported in previous studies (Sec.~\ref{result:gmc:f_coll}), our post-processing analysis reveals a significantly lower frequency by a factor of $4-5$ compared to these earlier results. This discrepancy is well predicted by our more stringent criteria for CCC identification, as embodied in criterion~\ref{ccc_cond2}.
It is important to emphasize that our approach differs from previous research, as exemplified by papers such as \citet{tasker_2009}, \citet{tasker_2011} and \citet{dobbs_2015}, which have predominantly examined how often clouds undergo ``mergers'' within a given time span. In contrast, our focus is on the occurrence of cloud-cloud ``collisions''; we exclude grazing contacts from collisions, as such mergers are unlikely to be important in triggering enhanced star formation. 
When we impose only the criterion~\ref{ccc_cond1} for CCC identification in the post-processing analysis, the collision frequencies are between $\sim40-50\,\gyr^{-1}$. 
The average collision frequency for this identification is $\bar{f}_\mathrm{ccc}\approx44.2\,\gyr^{-1}$, which is comparable to the previous studies.

We lastly show the differences in the values of $f_1$ and $f_2$ in Fig.~\ref{fig:collision_frac_otfppa}.
The values of $f_1$ in the post-processing analysis have a wide distribution ranging from $\sim 0.09$ to $\sim 0.88$, with the most frequent occurrences being between $\sim 0.4$ and $0.7$. In contrast, the values for the on-the-fly analysis are typically $\gtrsim 0.5$.
For $f_2$ it becomes less likely as the value increases and both analyses show the same trend. However, the post-processing analysis cannot reach the maximum value of $f_2=0.5$ achieved by the on-the-fly analysis.
Despite the pronounced differences in the distributions of $f_1$, there are relatively similar distributions for $f_2/f_1$, while the distributions for $f_1+f_2$ show significant differences.
The probability of $f_2/f_1$ in the post-processing decreases with increasing value, mirroring the trend seen in the on-the-fly analysis. However, the distribution of $f_2/f_1$ in the post-processing analysis is flatter than in the on-the-fly analysis, indicating that the ratio of $f_2$ to $f_1$ tends to become larger when a longer time interval is used to identify CCCs.

The post-processing analysis yields a broad distribution of $f_1+f_2$ spanning from $\sim 0.18$ to $\sim 0.93$. The median value, around $\sim 0.66$, is comparable to the minimum value in the on-the-fly analysis.
This outcome indicates that the post-processing analysis typically involves the contribution of more than two progenitor GMCs.
Given that the time interval of $1 \, \myr$ used in the post-processing analysis is too long and that as short as the hydrodynamical timestep is needed to study CCCs. 
If we use only the criterion~\ref{ccc_cond1} for CCC identification in the post-processing analysis, the probability density of $f_2/f_1$ becomes even flatter, while the distributions of $f_1$, $f_2$ and $f_1+f_2$ remain mostly unchanged.

\subsection{Cloud-cloud collisions and galactic structures}
\label{discussion:gal_morphology}

We have carried out simulations of isolated disk galaxies without imposed spiral arms or a bar, using a static potential to mimic a MW-like rotation curve. 
Using a static potential reduces the computational cost by removing the need for dark matter and stellar particles to shape galactic structures. 
However, the interaction between baryons and dark matter, which affects development of the galactic structures, is recognized \citep[e.g.][]{kim_2016}. 
Furthermore, to study star formation within dynamically evolving spiral arms or a bar, it is necessary to include stellar particles in the initial conditions, as opposed to relying solely on a static potential \citep[e.g.][]{pettitt_2017,iles_2022}.
In future work, we plan to use a live stellar disc to study the relationship between galaxy evolution and star formation in colliding clouds.

Structures within galaxies, such as spiral arms and bars, are thought to influence both star formation and cloud evolution \citep[e.g.][]{momose_2010,schinnerer_2017}.
\citet{dobbs_2015} found that a simulated galaxy with a spiral potential had a merger frequency about three times higher than in the absence of spirals. This finding highlights the enhanced role of CCCs in spiral galaxies compared to the context of this study.
Examination of the bar region by \citet{fujimoto_2014_b} revealed a higher collision speed, potentially explaining the subdued star formation observed in bars \citep[e.g.][]{momose_2010}. However, these investigations were based on post-processing analyses and lacked an integrated star formation model for CCCs.
Furthermore, \citet{finn_2019} observed the higher collision velocity ($>100 \, \mathrm{km}~\mathrm{s}^{-1}$) during the merger of Antennae galaxies, a phenomenon we cannot reproduce in our simulations.
Understanding how CCCs affect merging galaxies, expected sites of intense star formation, is important for both star formation and galaxy evolution. We now have the ability to self-consistently simulate triggered star formation within such galactic structures using our on-the-fly collision detection algorithm.
In our future work, we will investigate the relationship between CCCs, galactic star formation and environmental effects.
The CEERS project by JWST investigates the assembly of galaxies between the reionisation epoch and today by probing the build-up of their stellar mass and their morphological transformation \citep[][]{kartaltepe_2023}.
Our forthcoming simulations will contribute to the understanding of the outcomes of the CEERS project.

Although the majority of low-mass stars in the MW probably form in smaller clouds with masses below $10^4\,\Msun$, which our simulations cannot readily identify, or in non-colliding cloud environments, current observational evidence does not provide compelling evidence that high-mass star formation occurs primarily in such small clouds \citep[e.g.][]{tan_2014}.
In contrast, \citet{fukui_2021} argue that CCCs play a key role in facilitating high-mass star formation within the MW. In addition, CCCs are also thought to contribute to the formation of low-mass stars and cores \citep[e.g.][]{takahira_2018}.
It is worth noting that the study presented in this paper is the first step in exploring the relationship between CCCs and star formation within galaxy-scale simulations. 
In order to develop a comprehensive understanding, especially with regard to low-mass stars, further in-depth investigations will be essential in the future.

\section{conclusion}
\label{conclusion}

We have studied star formation triggered by CCCs and GMC properties using simulations of an isolated disk galaxy.
To account for the triggered star formation in galaxy simulations, we have newly developed the on-the-fly algorithm that identifies CCCs at each timestep and the subgrid model of star formation by CCCs.
We have performed the two simulations:
    the \textit{Const} simulation employs the constant star formation efficiency per free-fall time parameter and the \textit{Coll} simulation turns on the star formation model of CCCs.
This work is the first to self-consistently consider star formation induced by the collision of GMCs and the implications of such star formation for galaxy evolution using galaxy simulations.  

We have found that the enhanced star formation within CCCs influences several aspects of the simulated galaxy. 
It is therefore essential to incorporate collision-induced star formation into our models in a self-consistent manner.
There are noticeable differences between the two simulations in the SFRs, the PDFs of the GMC masses, the KS relations, and the PDFs of the star formation efficiencies within the GMCs. These differences are due to the increased star formation activity within colliding GMCs applied to the Coll simulation. Meanwhile, other properties such as the collision velocity, the progenitor mass ratios of the colliding clouds, and the collision frequency show minimal dependence on the chosen star formation model for the colliding GMCs. 

As we showed in Sec.~\ref{discussion:ccc_identification}, CCC identification by post-processing with $1~\myr$ interval snapshots underestimates the collision frequency by a factor of $\sim 8$ compared with the on-the-fly identification. 
The longer interval also makes most of the collisions multiple, while most of the collisions identified by the on-the-fly approach are two-body collisions. 
The use of post-processing methods to identify CCCs, hence, makes it impractical to obtain accurate statistics on CCCs, including the frequency of CCC occurrences and collision mass ratios.

Our simulations are unable to replicate the exceptionally high collision speeds ($\varv_\mathrm{coll} > 100~\kms$) observed in Antennae galaxies \citep{finn_2019}. 
Our forthcoming focus entails studying merging galaxies to explore how CCC properties vary across diverse environments.

Our CCC-triggered star formation model relies solely on the collision velocity, inspired by the results of \citet{takahira_2018}. However, their simulations involve GMC collisions with mass ratios exceeding our typical value of $< 0.1$. As such low mass ratio CCC outcomes remain unexplored, our model may need to be revised. \citet{sakre_2021} demonstrated the effect of magnetic field strength on core formation in CCCs, while \citet{sakre_2023} highlighted the collision axis column density and velocity. Thus, an improved CCC-induced star formation model can take these factors into account.

\section*{Acknowledgements}
We thank the anonymous referee for their helpful comments which improved this manuscript.
Numerical computation and analyses were carried out on Cray XC50 and analysis servers at Center for Computational Astrophysics, National Astronomical Observatory of Japan.
Numerical calculations were performed in part using Oakforest-PACS at the CCS, University of Tsukuba and the computer resource offered under the category of General Project by Research Institute for Information Technology, Kyushu University.
SH acknowledges the financial support of JST SPRING, Grant Number JPMJSP2119.
TO acknowledges the financial support of MEXT/JSPS KAKENHI Grant 18H04333, 19H01931, and 20H05861.
This work is supported by MEXT as ``Program for Promoting Researches on the Supercomputer Fugaku'' (Toward a unified view of the universe: from large scale structures to planets).
We thank the developers of the following Python analysis tools: matplotlib \citep[][]{hunter_2007_matplotlib}, NumPy \citep[][]{vanderwalt_2011_numpy}, and yt \citep[][]{turk_2011_yt}.

\section*{Data availability}
The data underlying this article can be shared on a reasonable request to the corresponding author.











\bsp	
\label{lastpage}
\end{document}